\documentclass[12pt]{article}
\usepackage{epsf,epsfig,here,cite}
\newcommand{\fone}{\centerline{\bf Fig.}}
\newcommand{\ftwo}{\centerline{\bf Fig.}}
\newcommand{\fthree}{\centerline{\bf Fig.}}
\newcommand{\ffour}{\centerline{\bf Fig.}}
\newcommand{\ffive}{\centerline{\bf Fig.}}
\newcommand{\fsix}{\centerline{\bf Fig.}}
\newcommand{\fseven}{\centerline{\bf Fig.}}
\ifx\nofigs\undefined
\renewcommand{\fone}{\centerline{
\epsfig{file=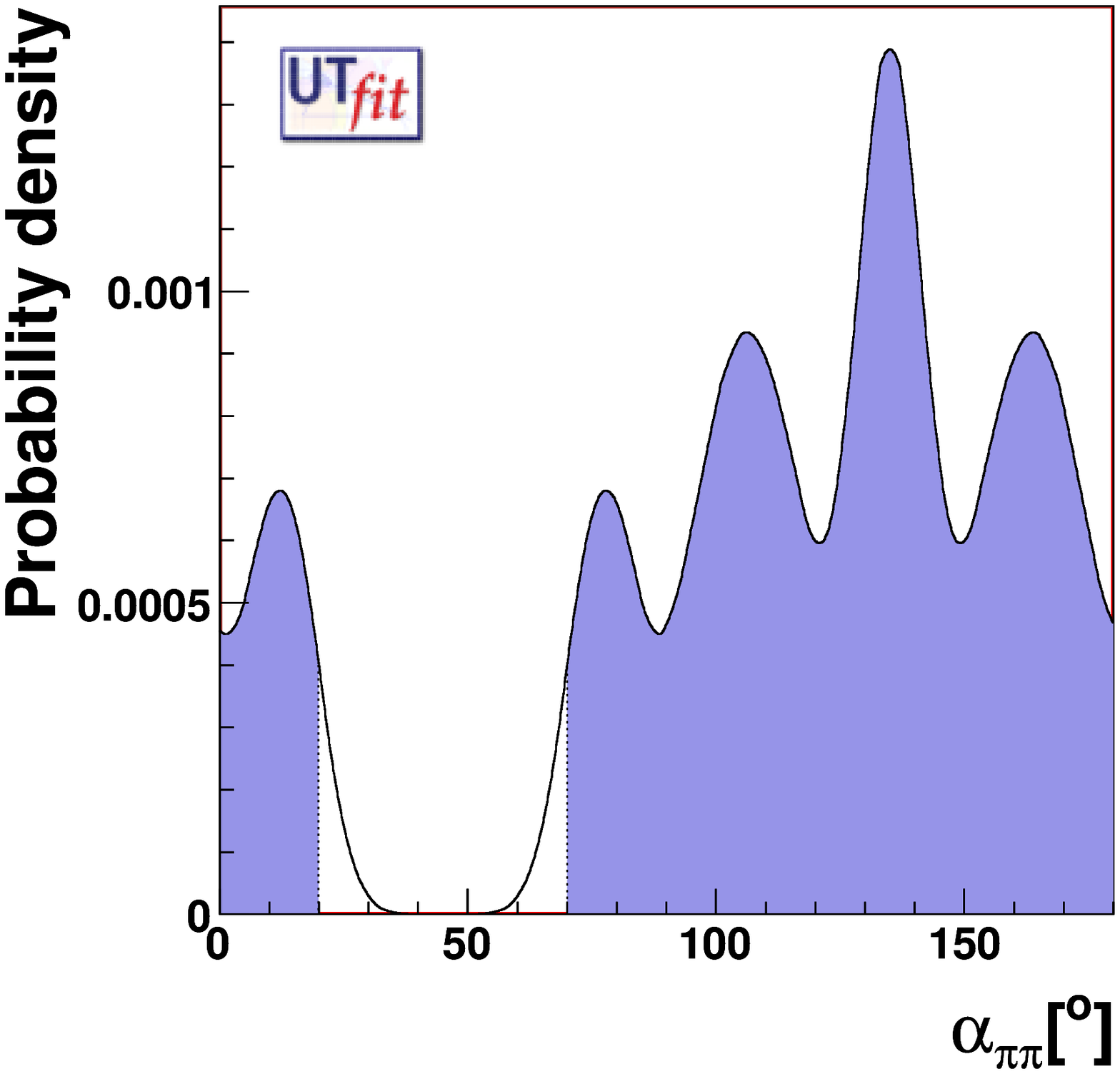,width=8cm}
\epsfig{file=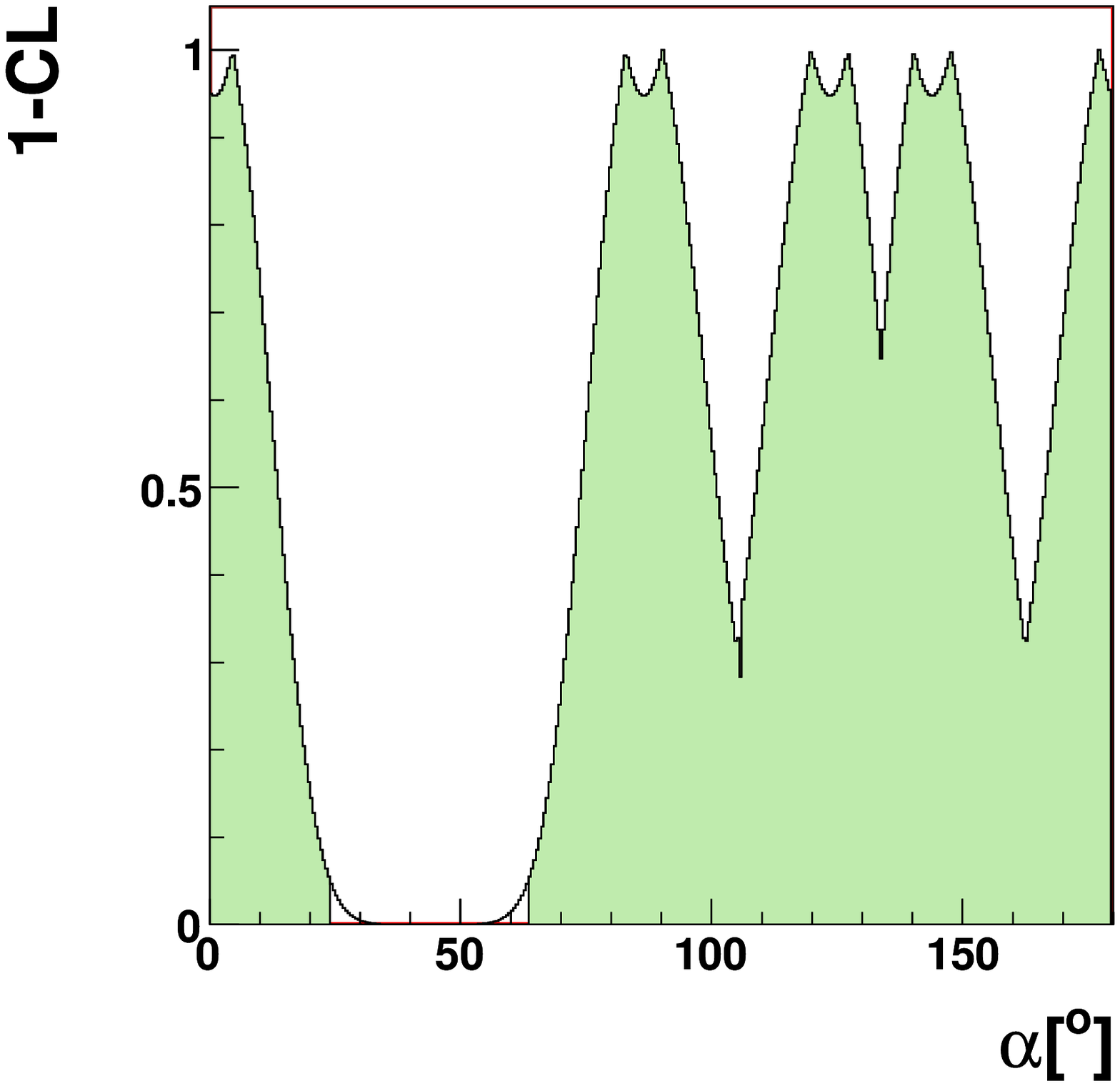,width=8cm}}}

\renewcommand{\ftwo}{\centerline{
\epsfig{file=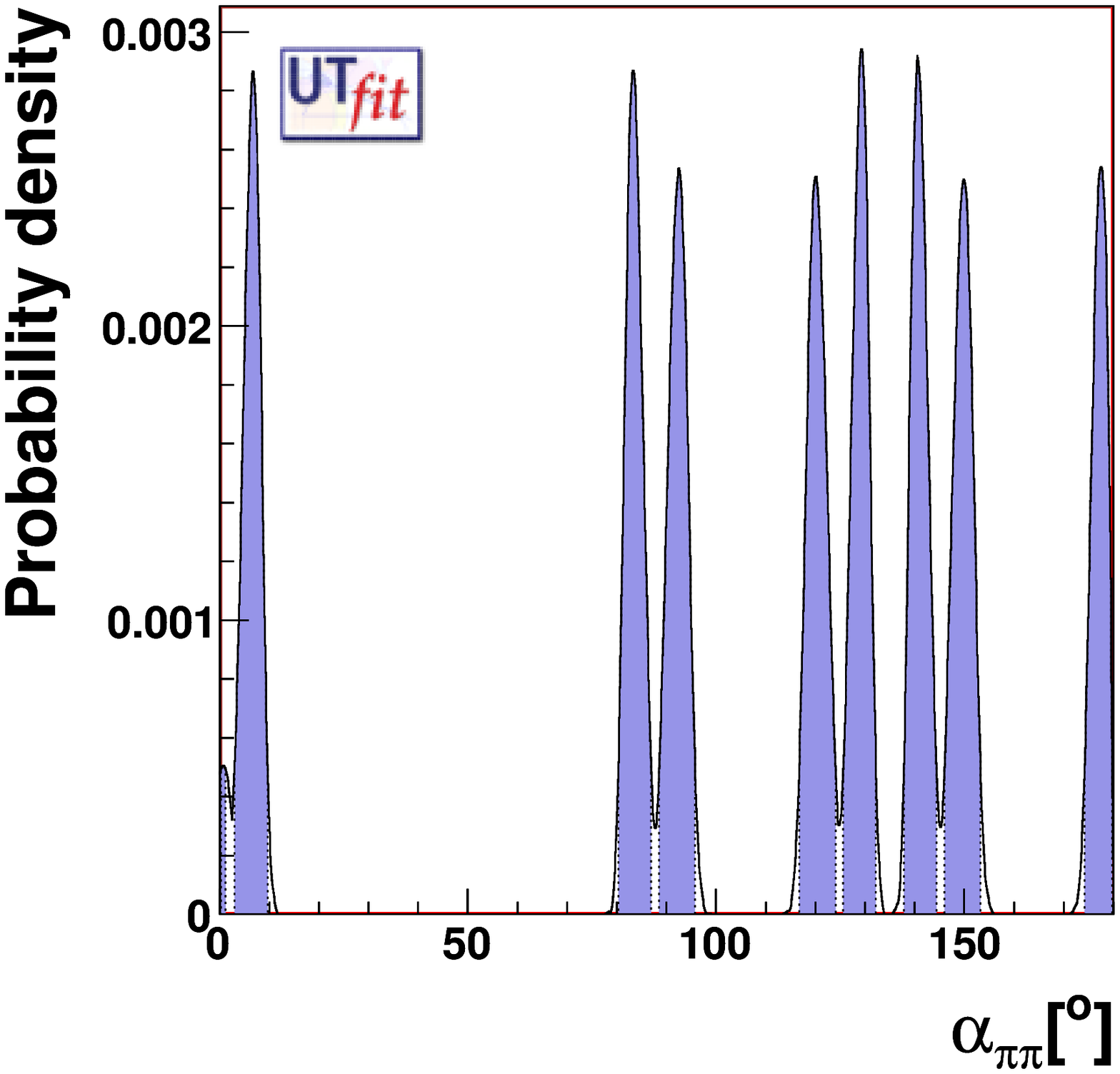,width=8cm}
\epsfig{file=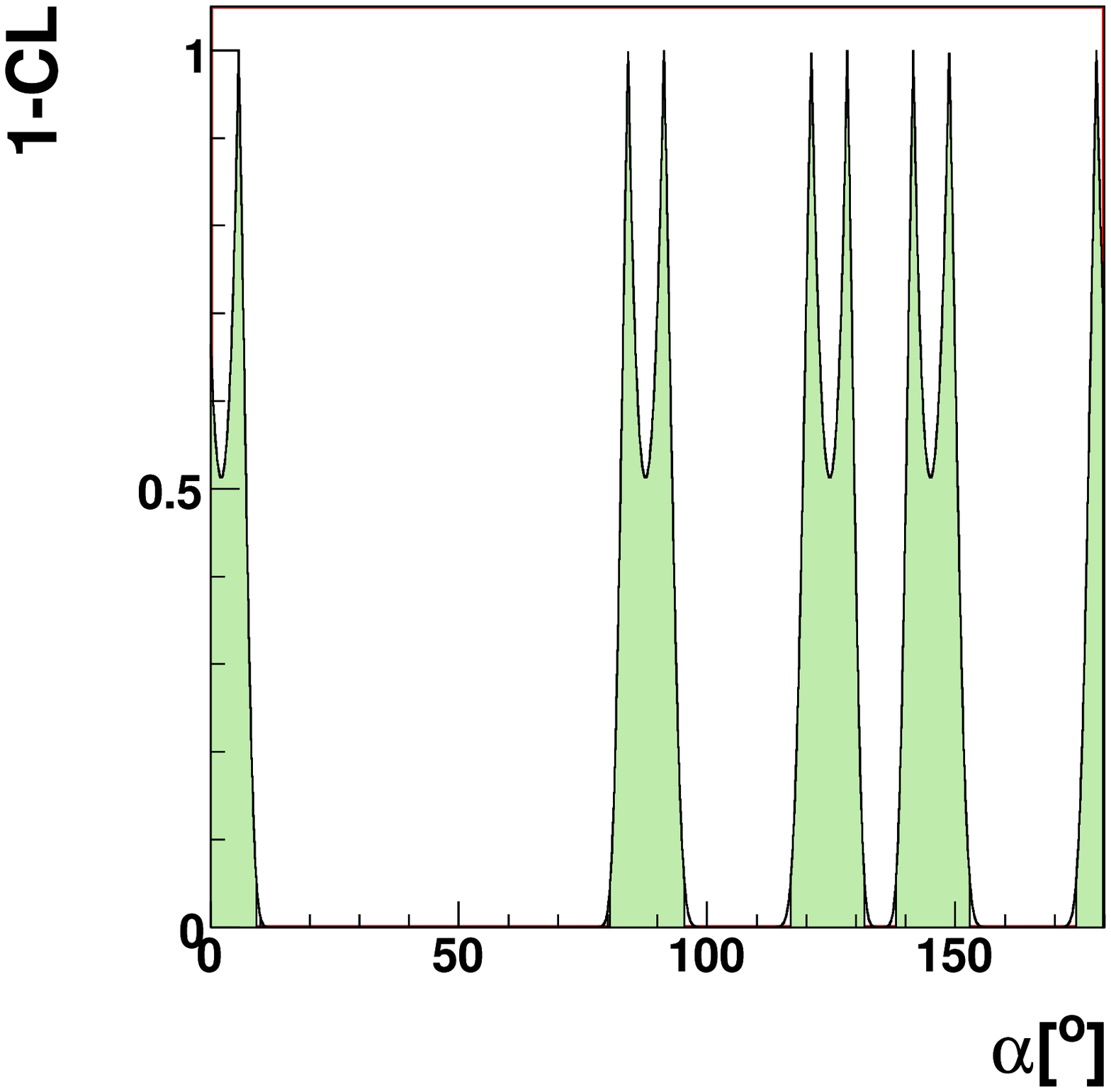,width=8cm}}}

\renewcommand{\fthree}{\centerline{
\epsfig{file=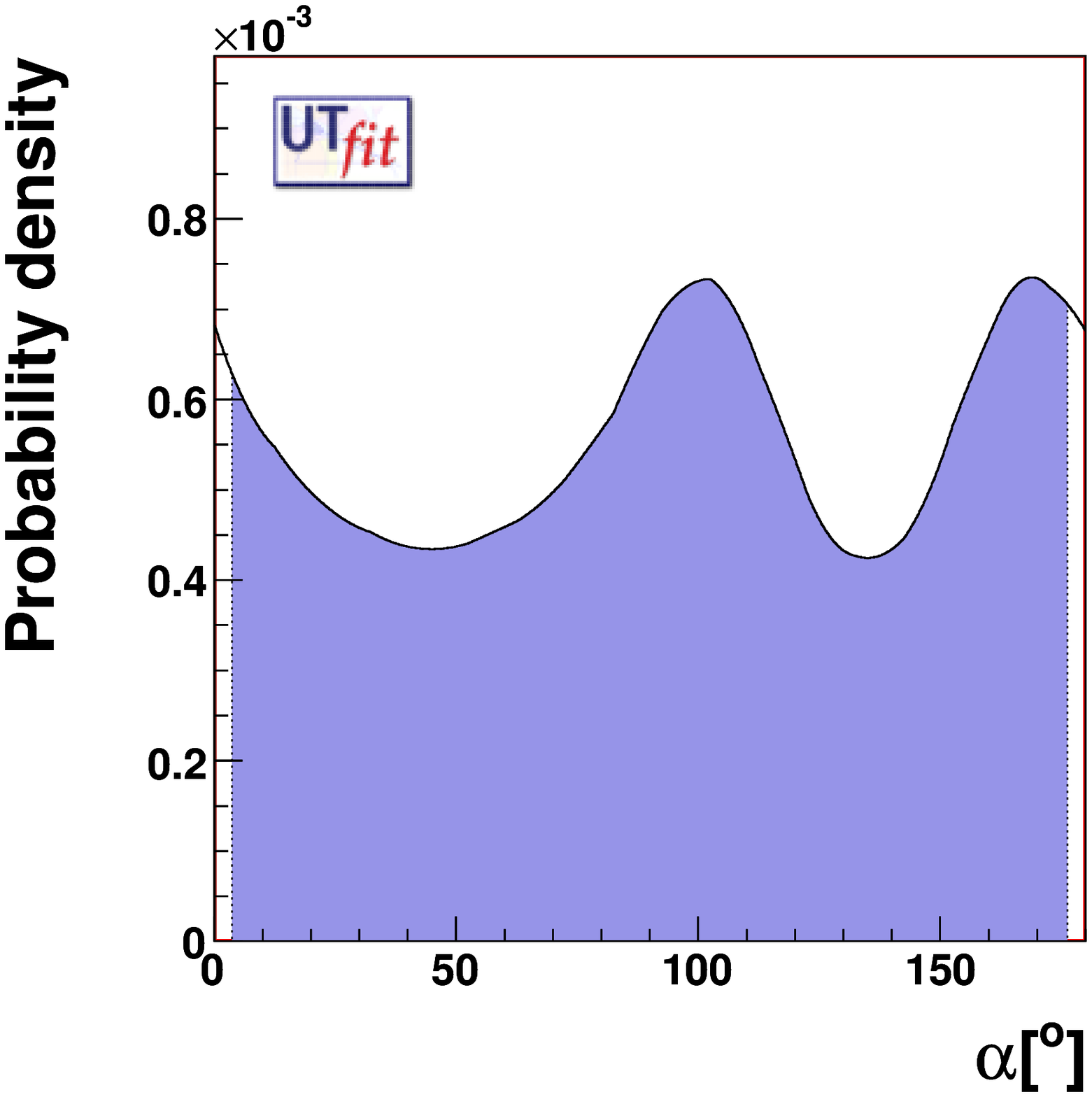,width=10cm}}}

\renewcommand{\ffive}{\centerline{
\epsfig{file=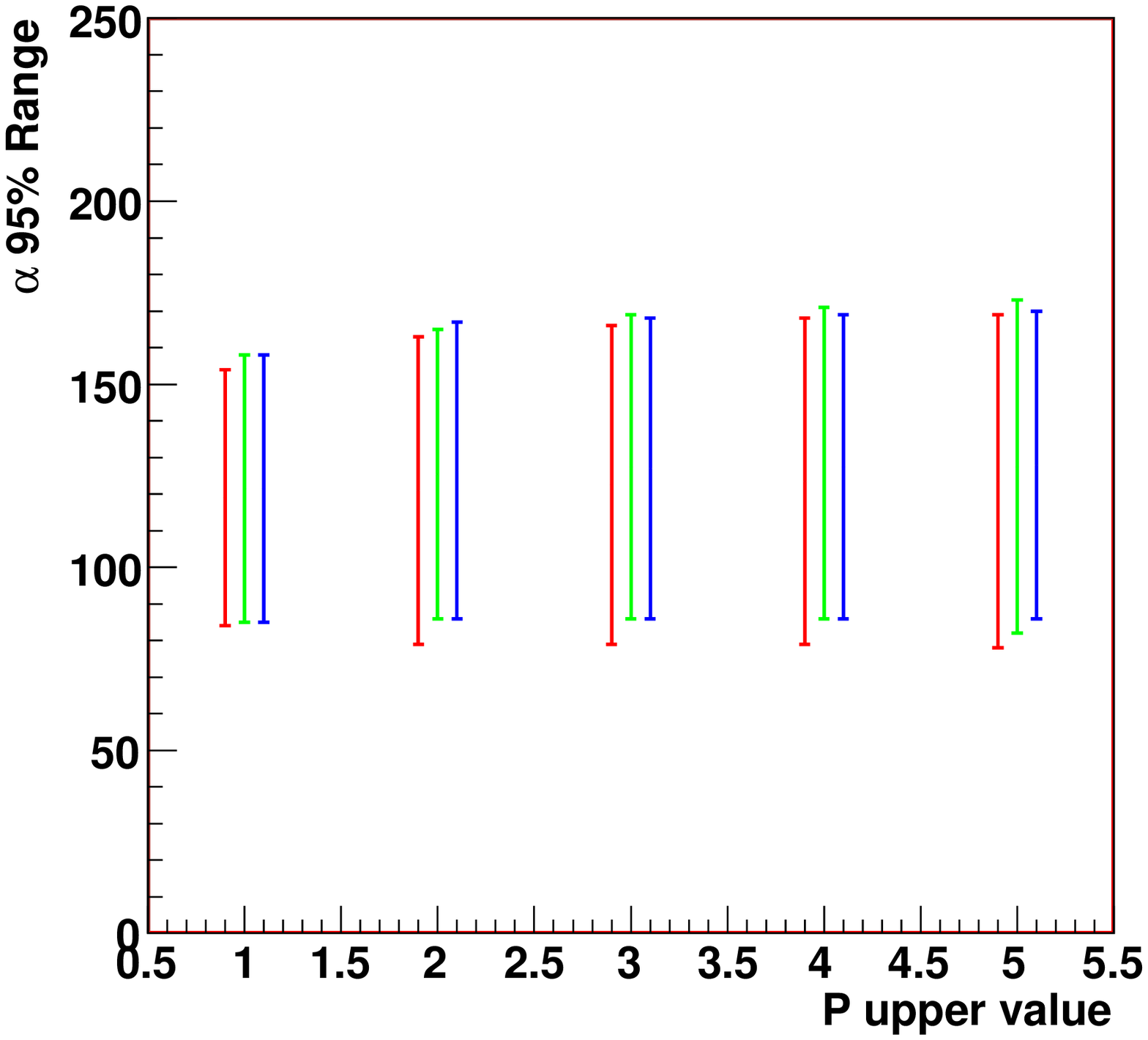,width=10cm}
}}

\renewcommand{\ffour}{\centerline{
\epsfig{file=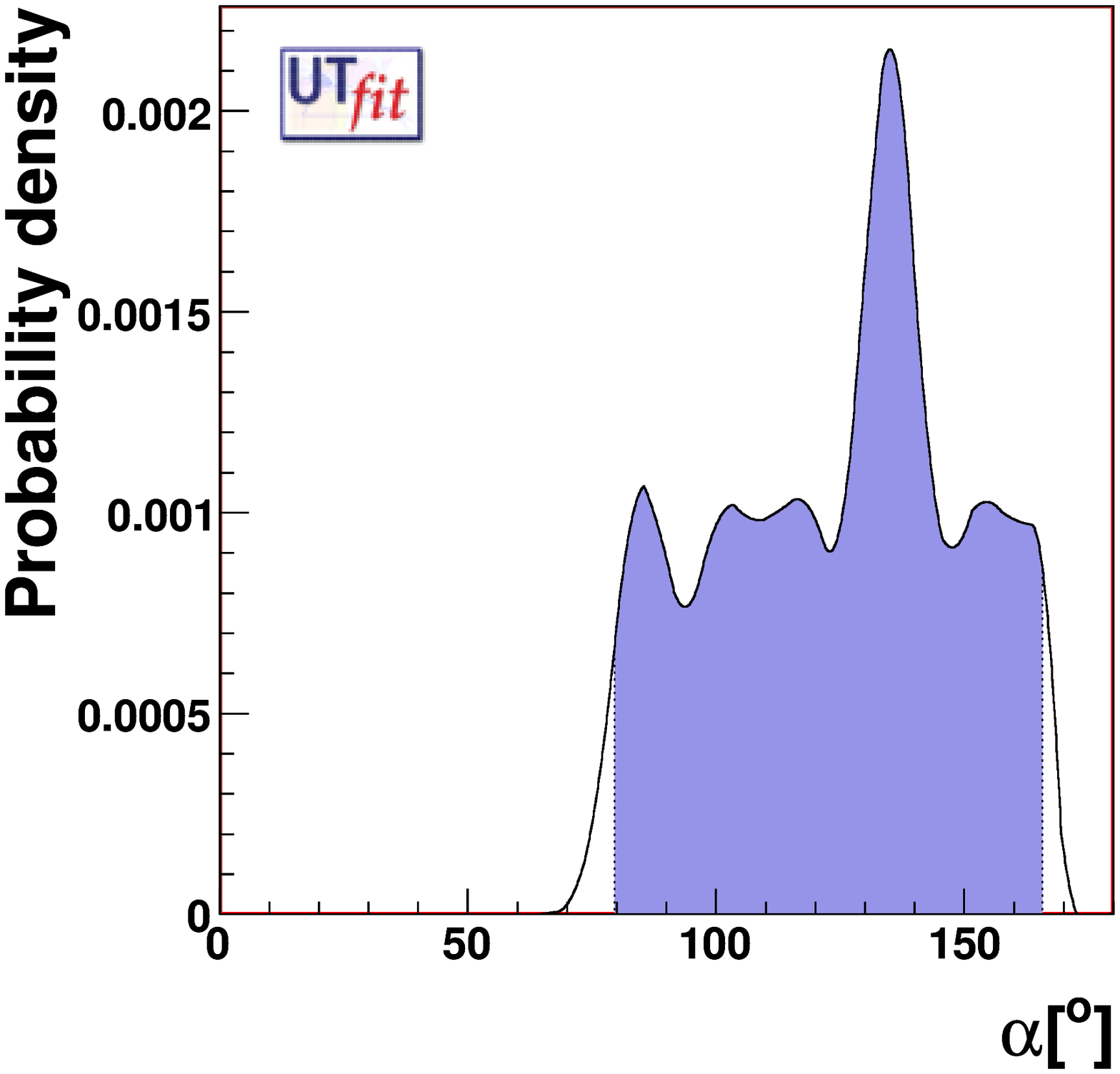,width=5cm}
\epsfig{file=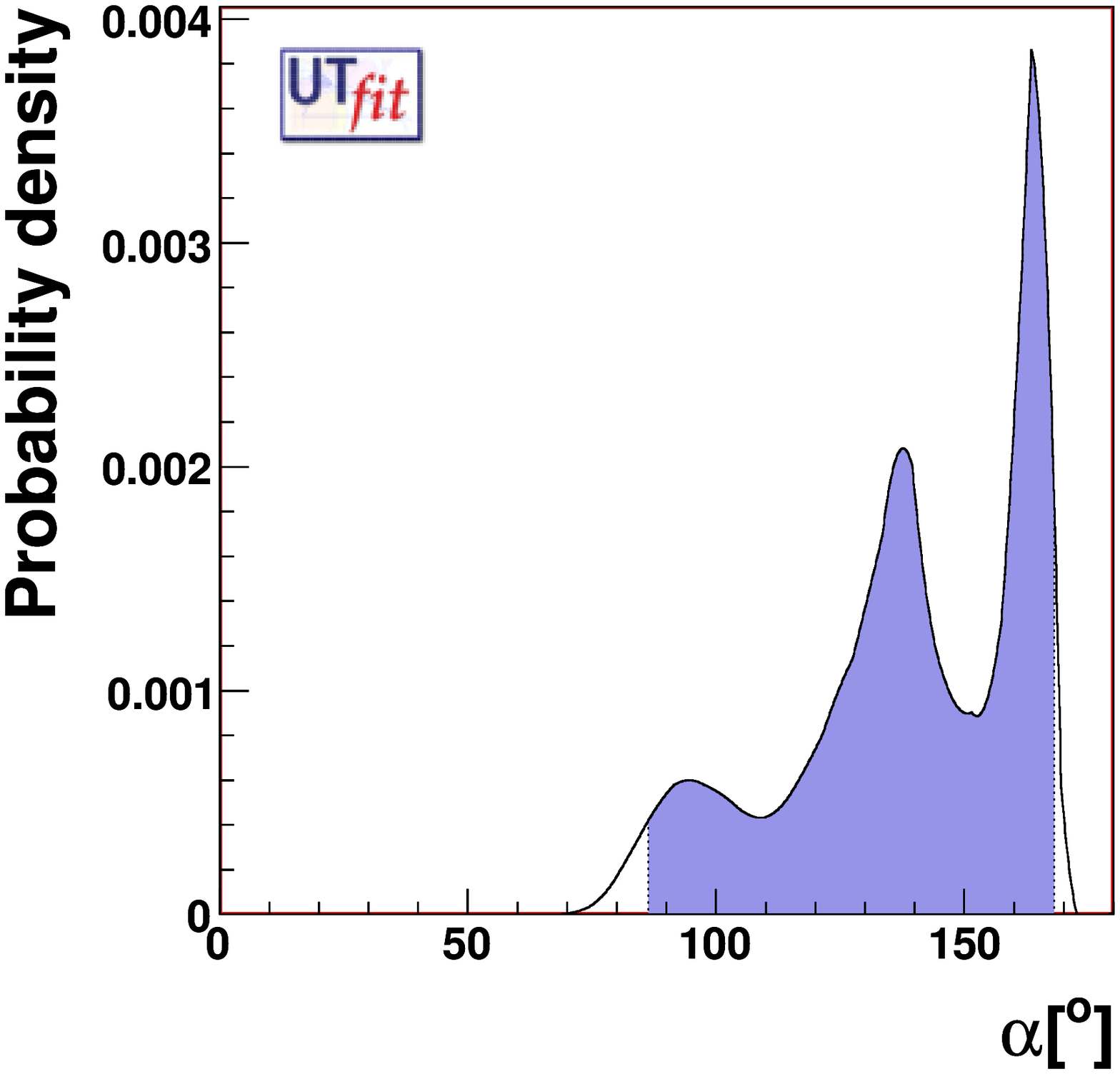,width=5cm}
\epsfig{file=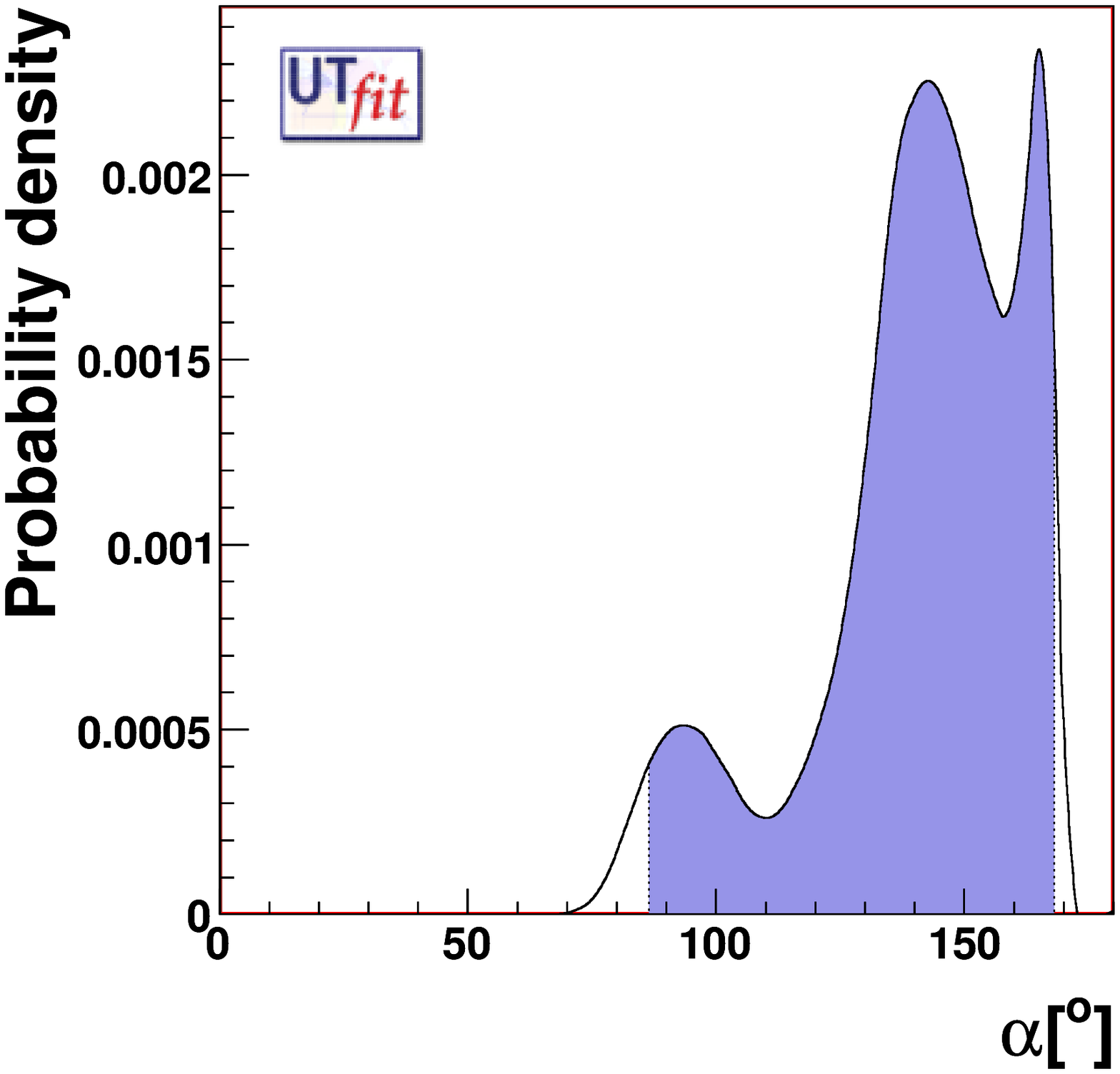,width=5cm}}}

\renewcommand{\fsix}{\centerline{
\epsfig{file=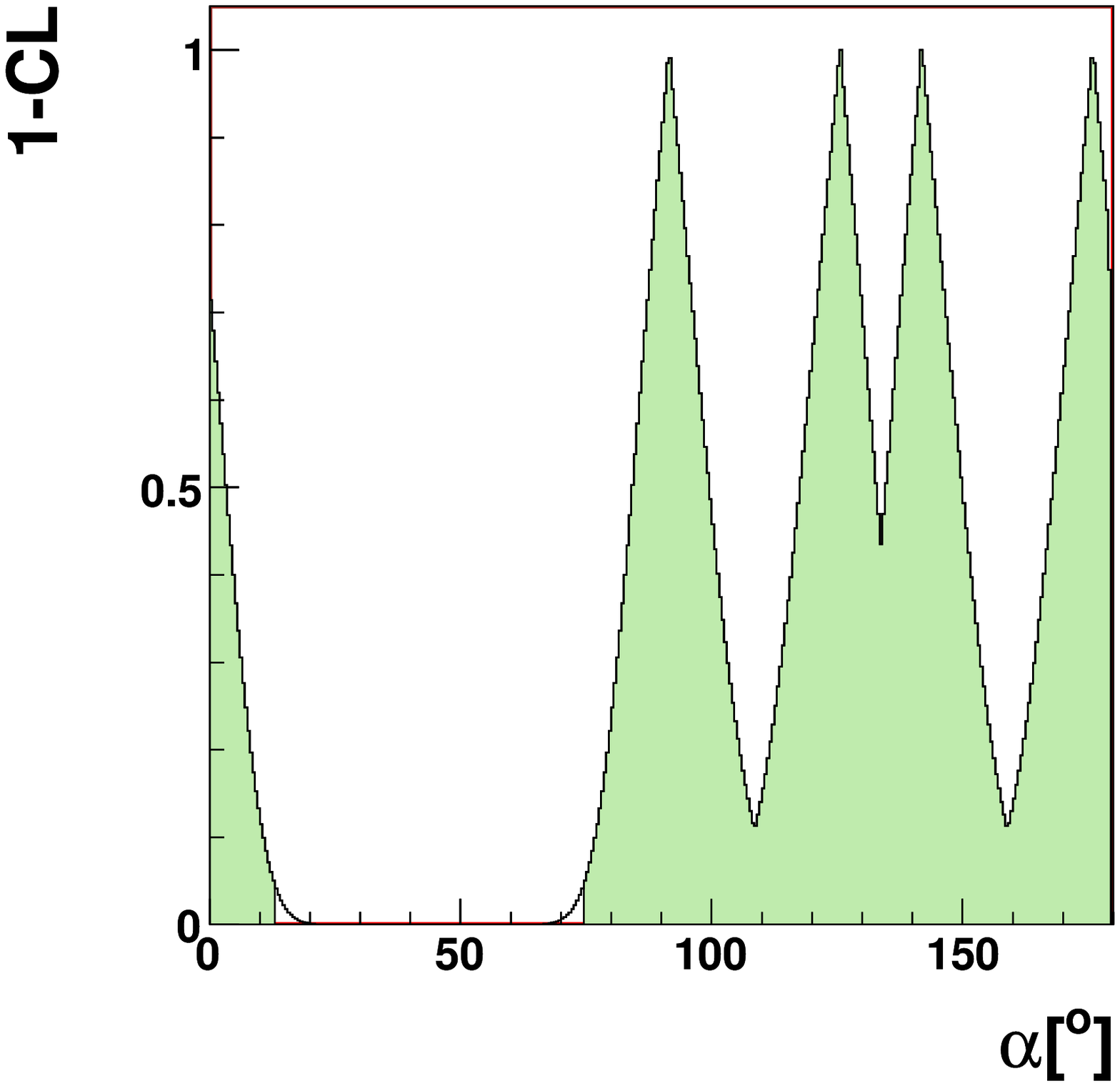,width=10cm}}}

\renewcommand{\fseven}{\begin{center}
\epsfig{file=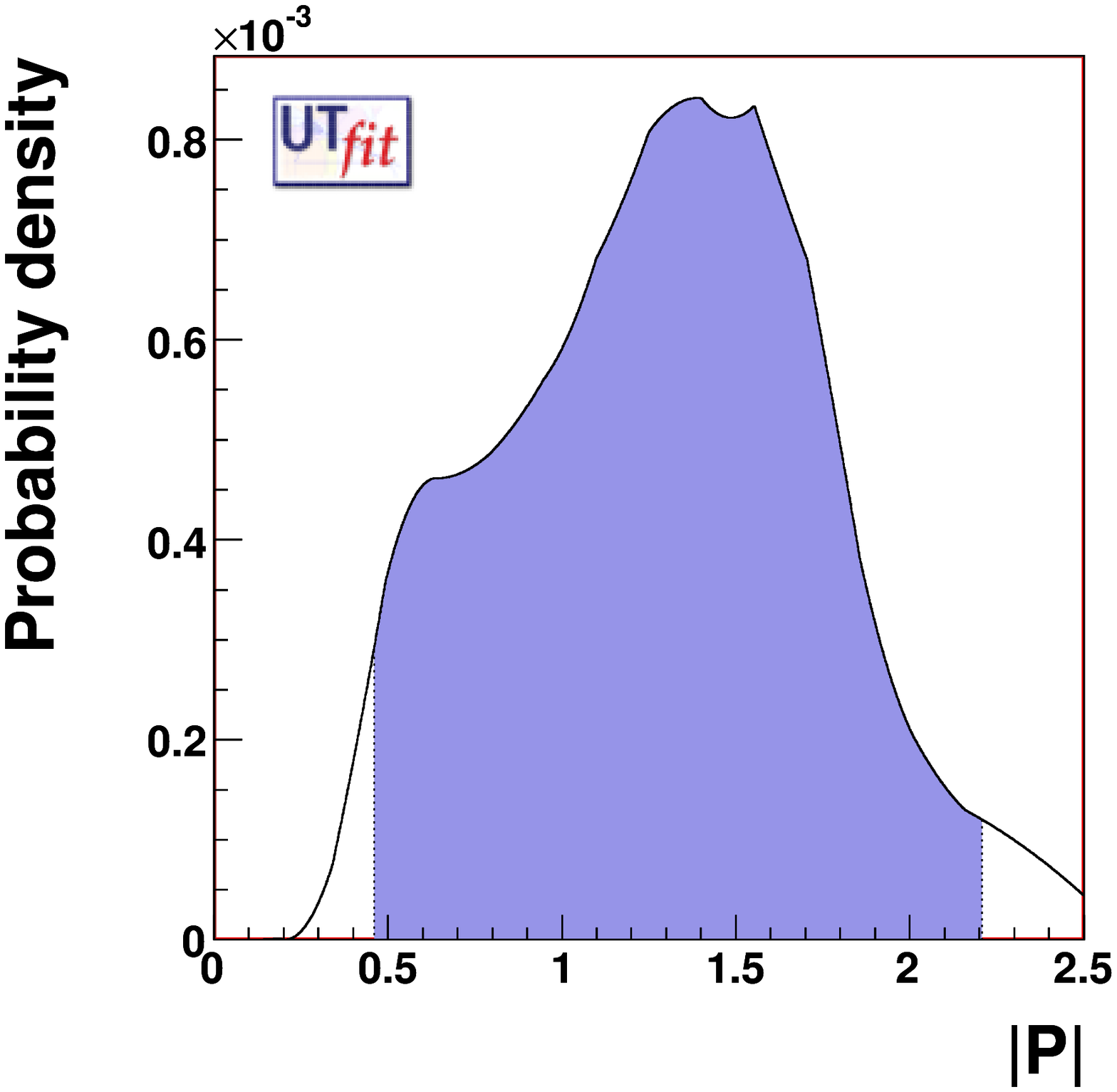,width=5cm}
\epsfig{file=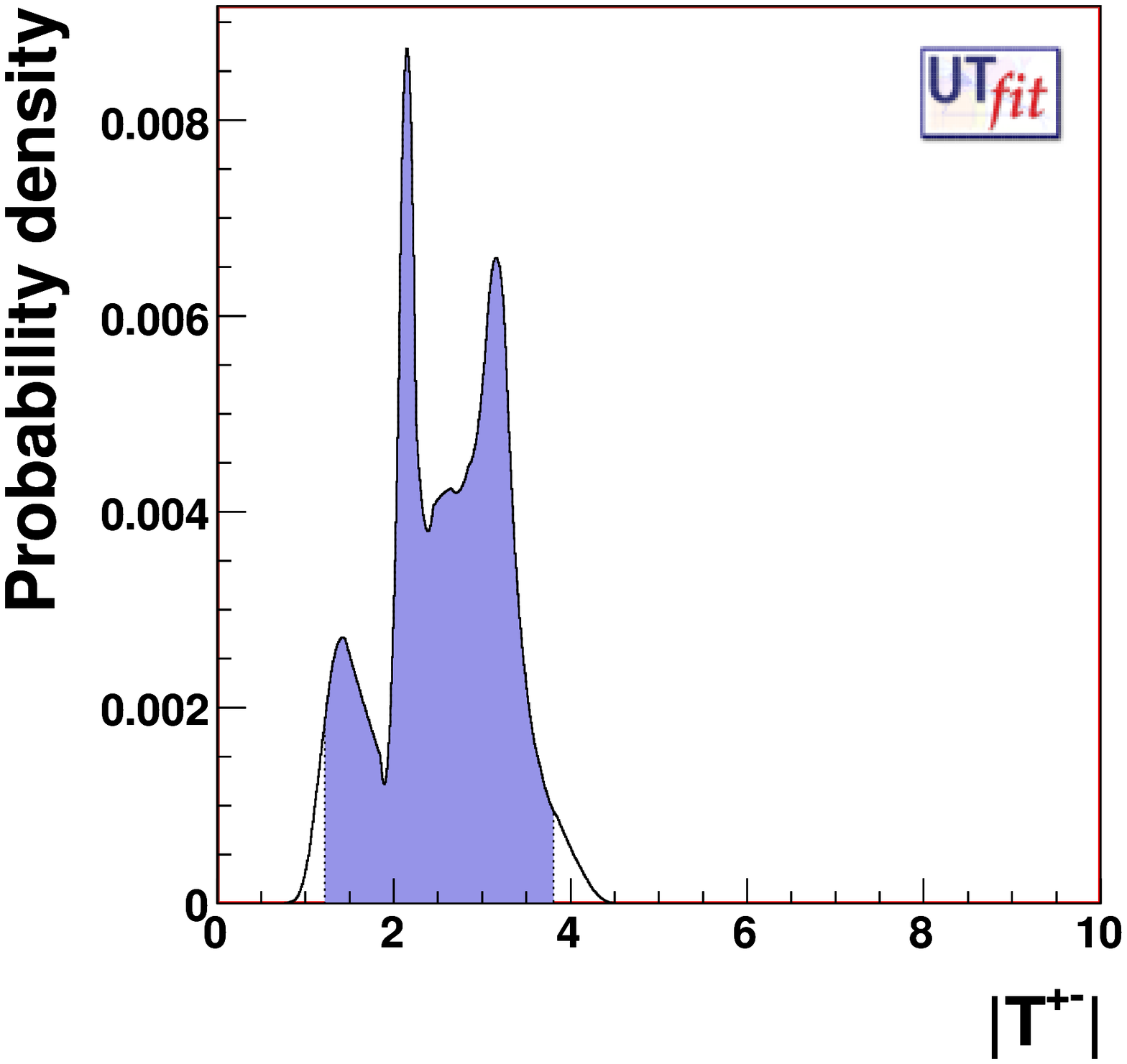,width=5cm}
\epsfig{file=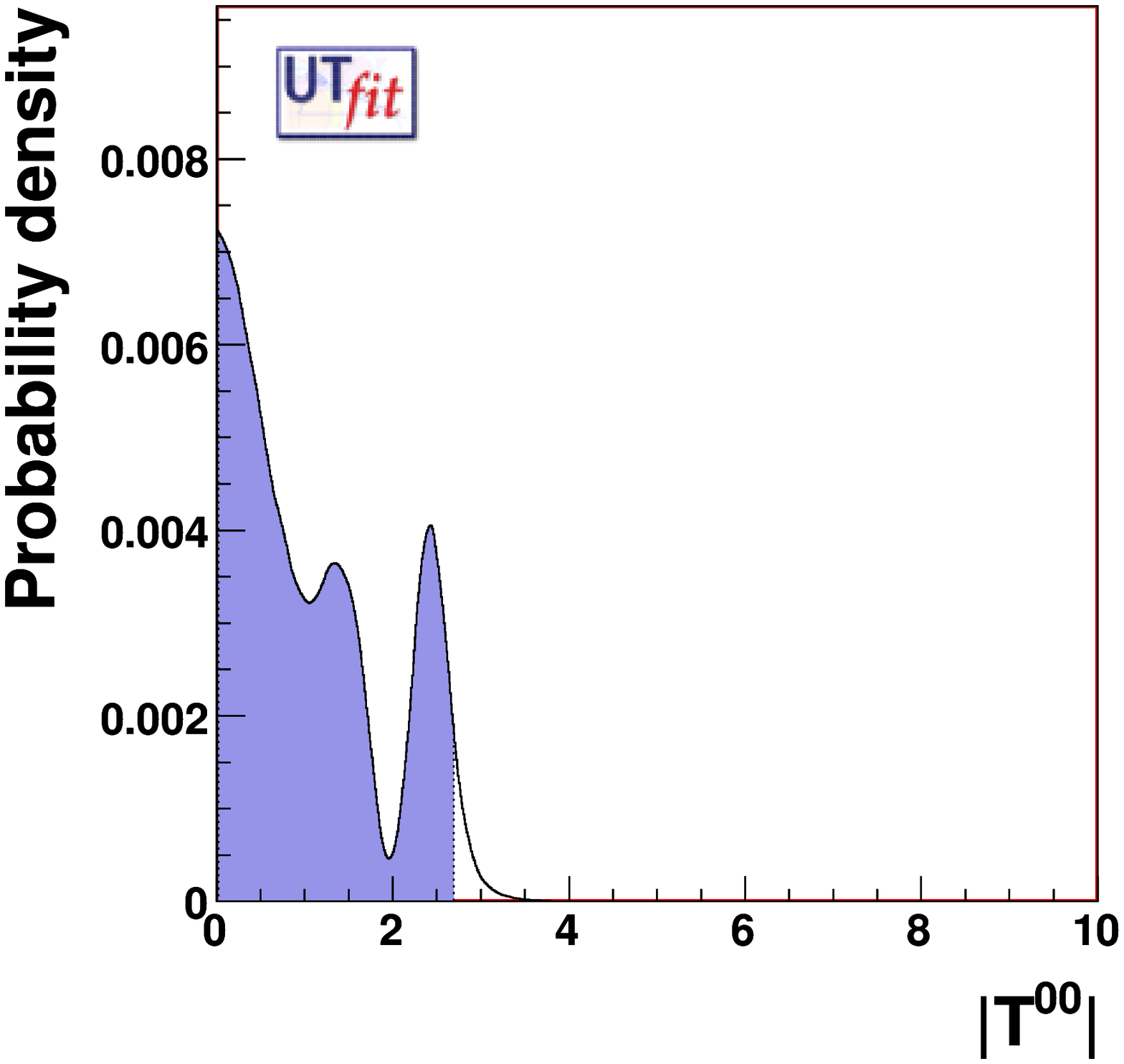,width=5cm}\\
\epsfig{file=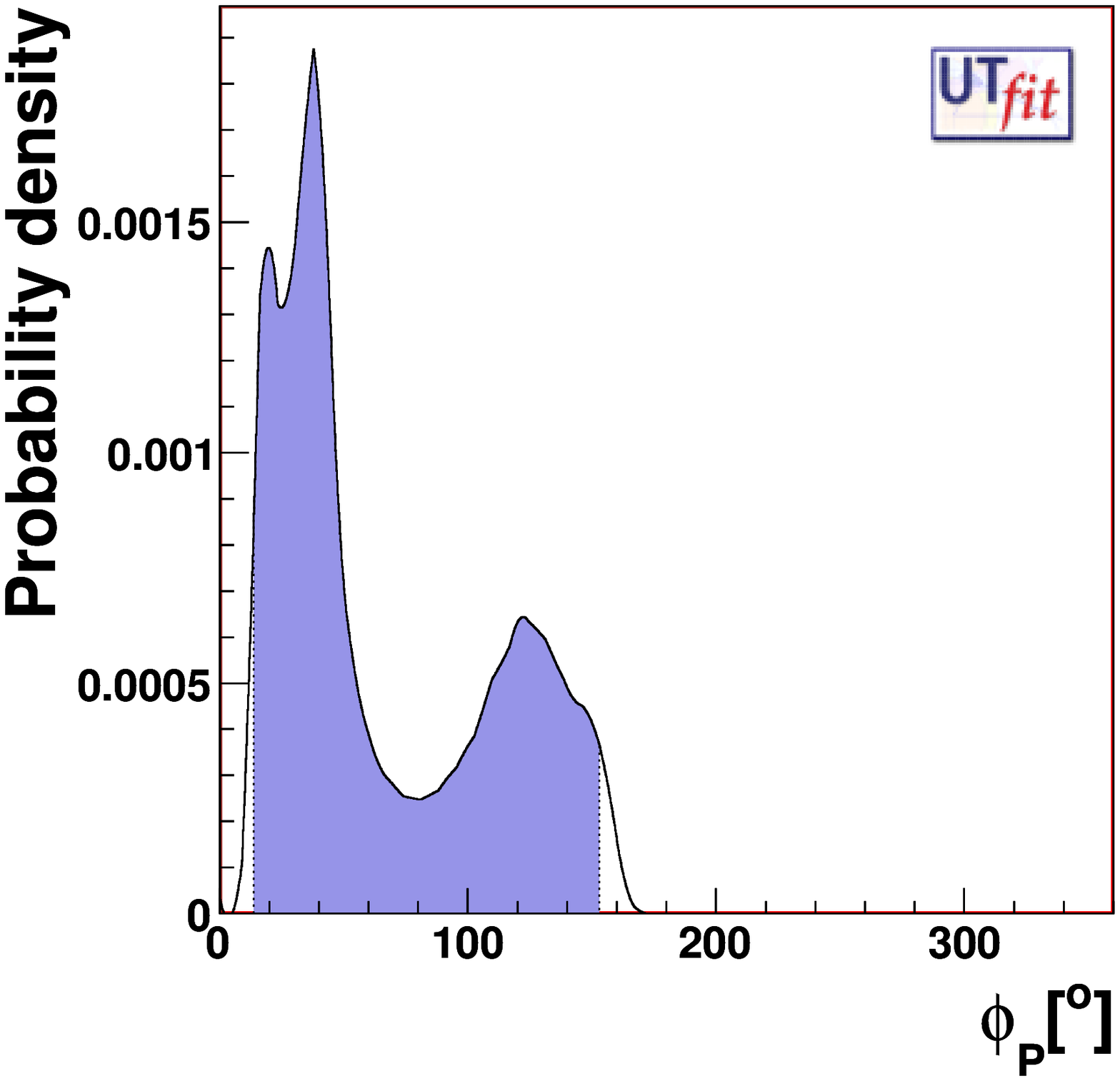,width=5cm}
\epsfig{file=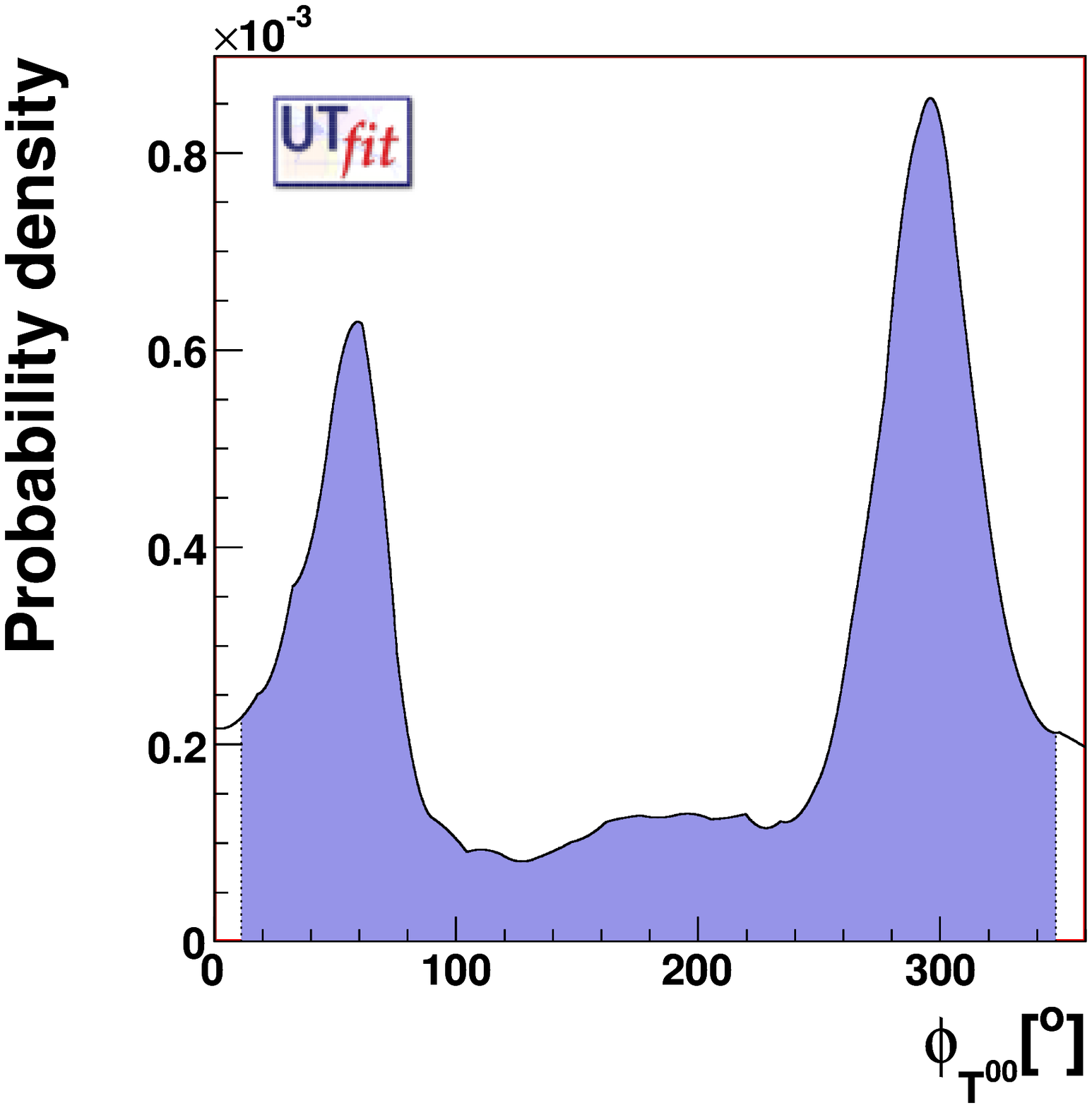,width=5cm}\\
\epsfig{file=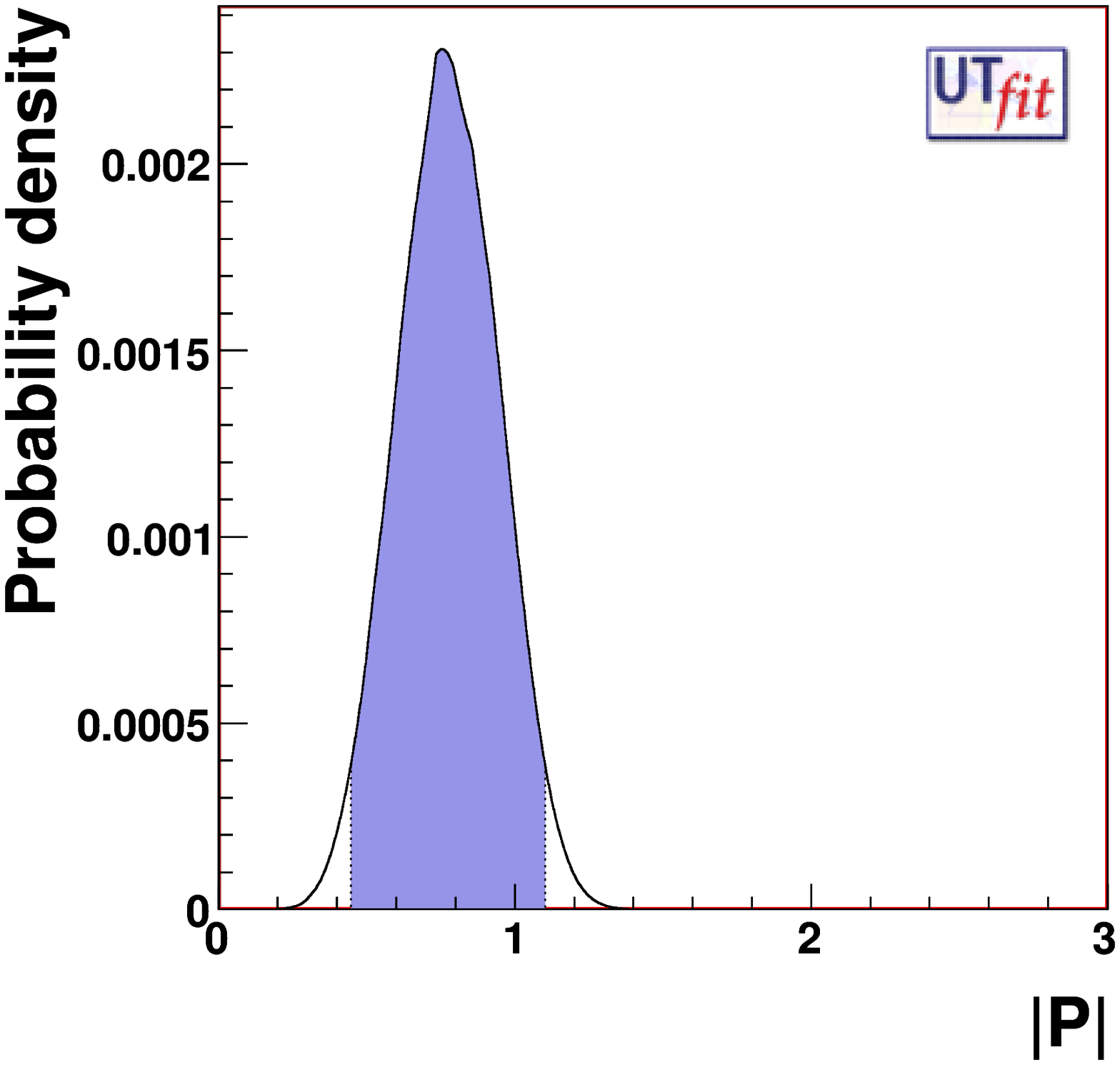,width=5cm}
\epsfig{file=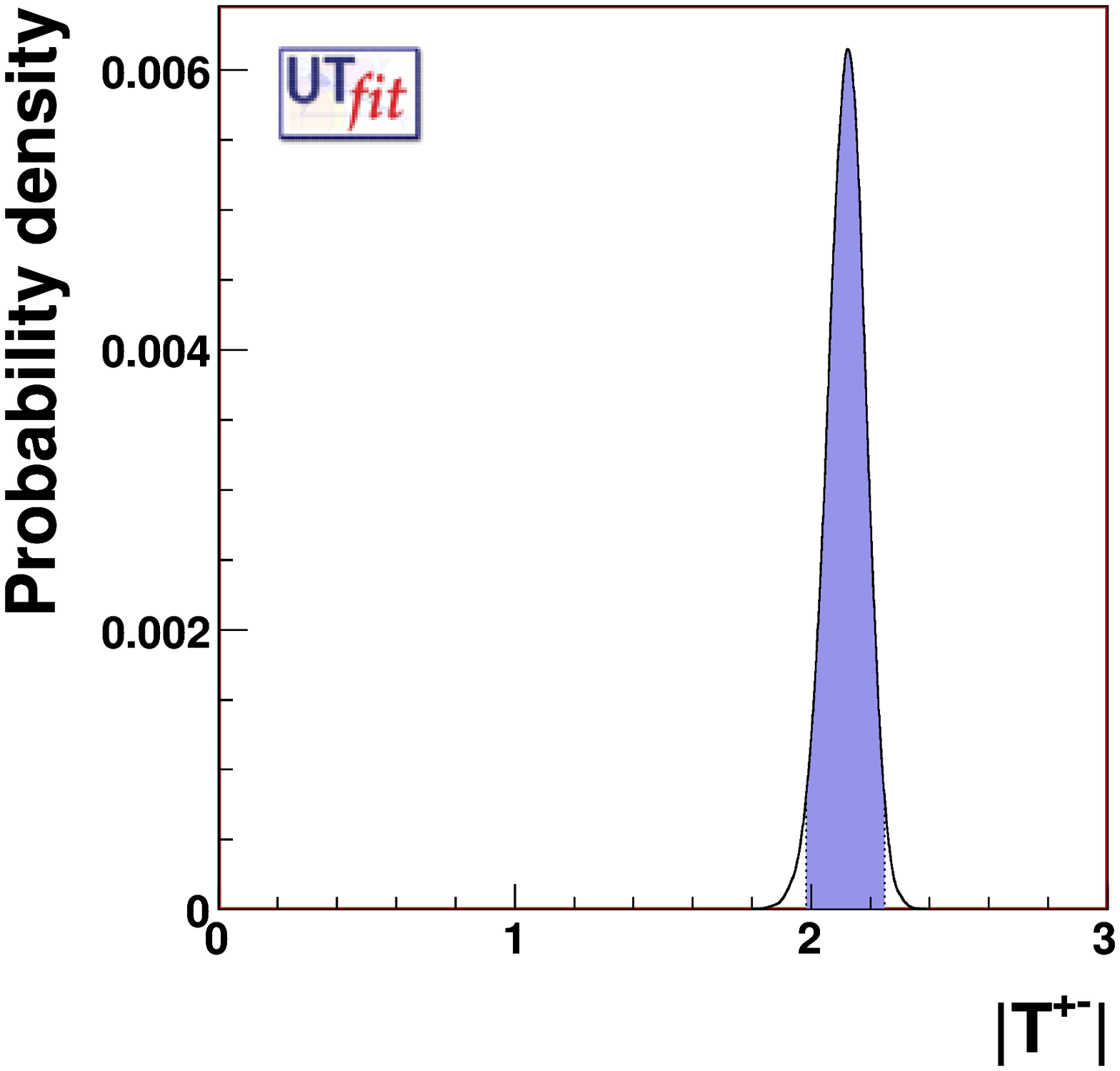,width=5cm}
\epsfig{file=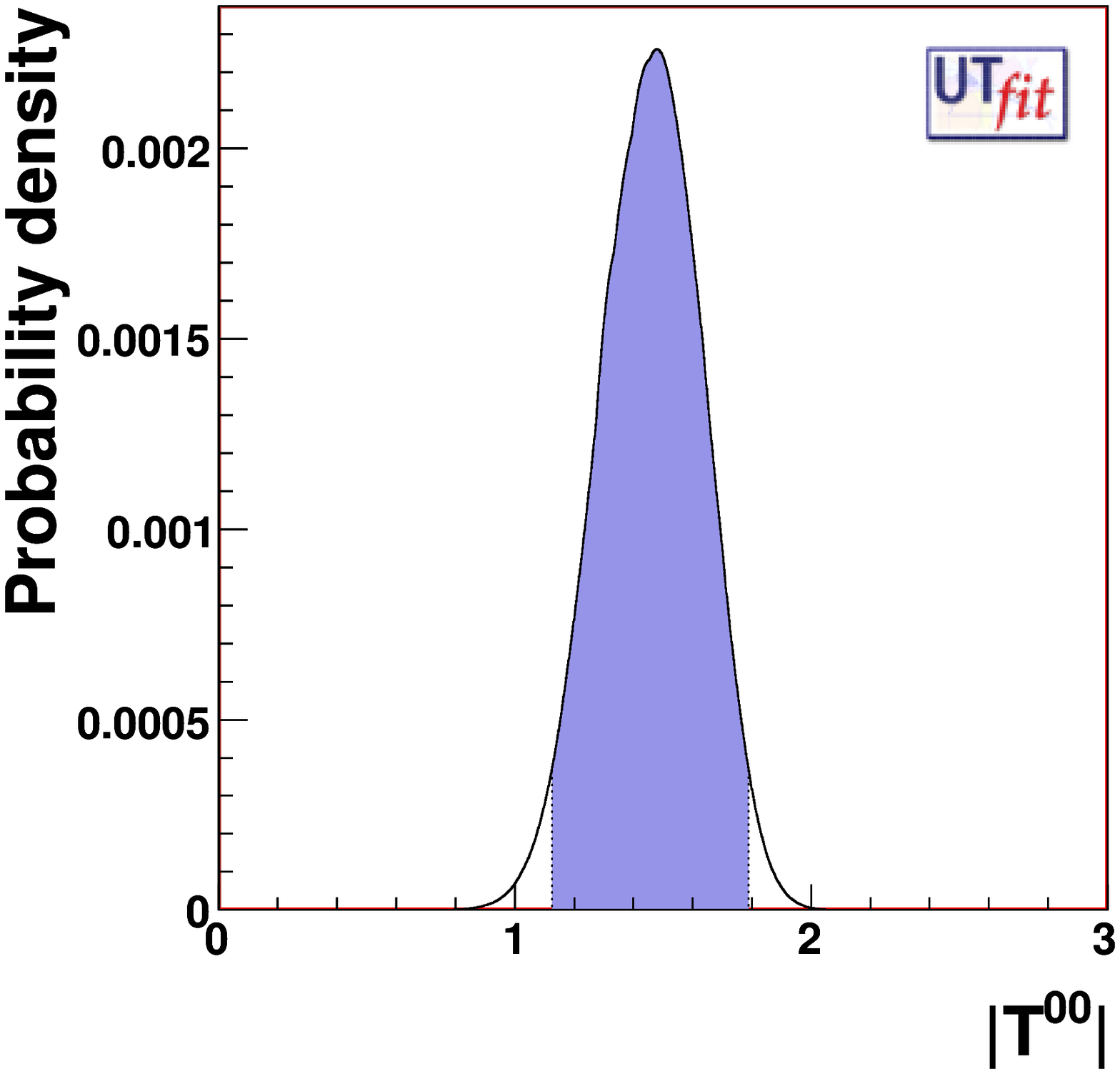,width=5cm}\\
\epsfig{file=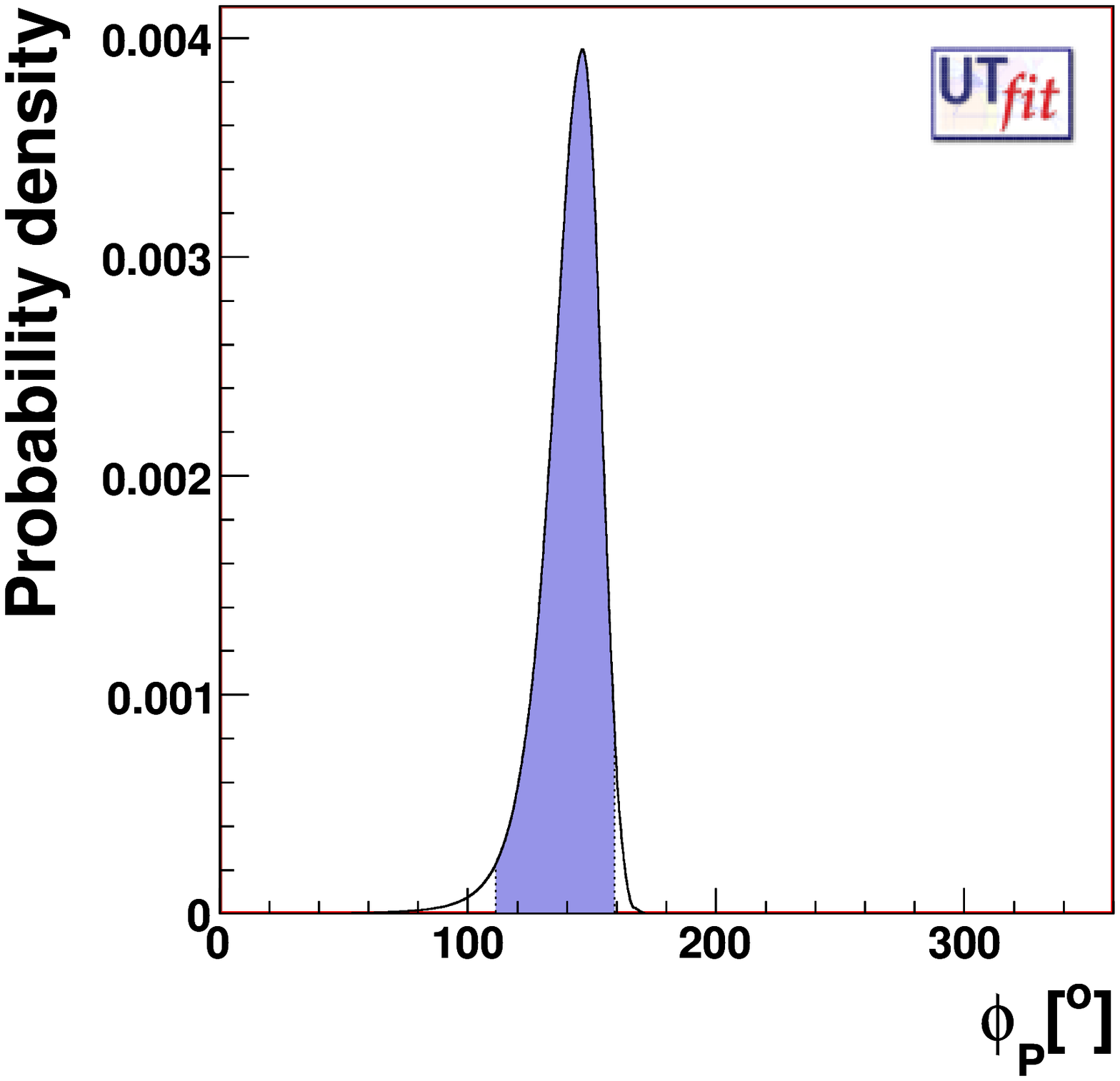,width=5cm}
\epsfig{file=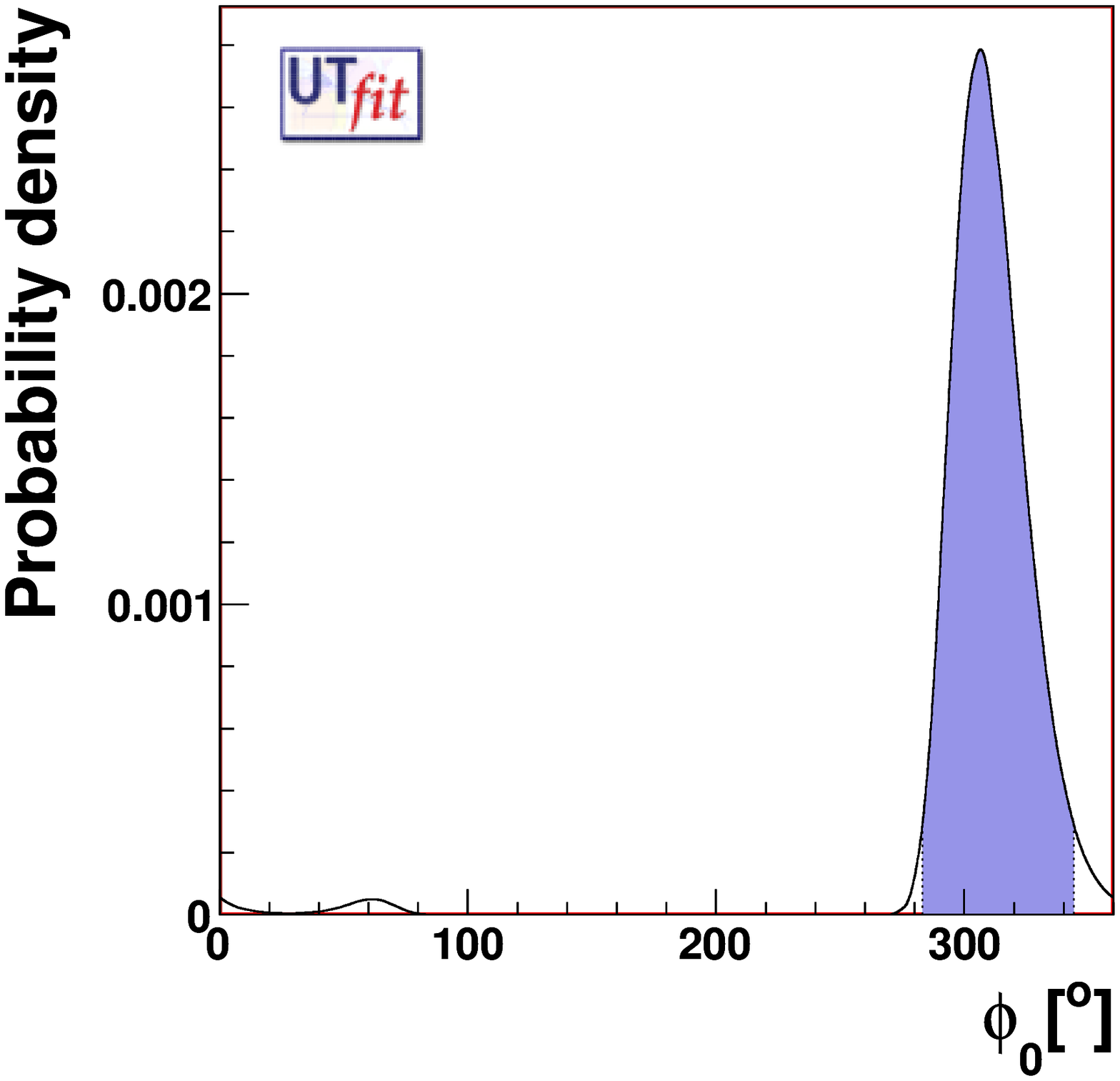,width=5cm}
\end{center}}

\fi

\newcommand{\beq}{\begin{equation}}
\newcommand{\eeq}{\end{equation}}
\newcommand{\beqn}{\begin{eqnarray}}
\newcommand{\eeqn}{\end{eqnarray}}
\newcommand{\bea}{\begin{eqnarray}}
\newcommand{\eea}{\end{eqnarray}}
\newcommand{\nn}{\nonumber}

\def\utfit{{\bf{U}}\kern-.24em{\bf{T}}\kern-.21em{\it{fit}}\@}
\def\utangles{{\bf{U}}\kern-.24em{\bf{T}}\kern-.21em{\it{angles}}\@}
\def\utlattice{{\bf{U}}\kern-.24em{\bf{T}}\kern-.21em{\it{lattice}}\@}

\begin{document}
\pagestyle{empty}
\pagenumbering{arabic}
\vskip  1.5 cm
\begin{center}
  \begin{LARGE}
    \textbf{Improved Determination of the  CKM Angle \\ $\alpha$   from $B \to \pi \pi$ decays} \\
  \end{LARGE}
\end{center}

\vspace*{-0.2cm}
\begin{figure}[htb!]
 \begin{center}
   \epsfig{figure=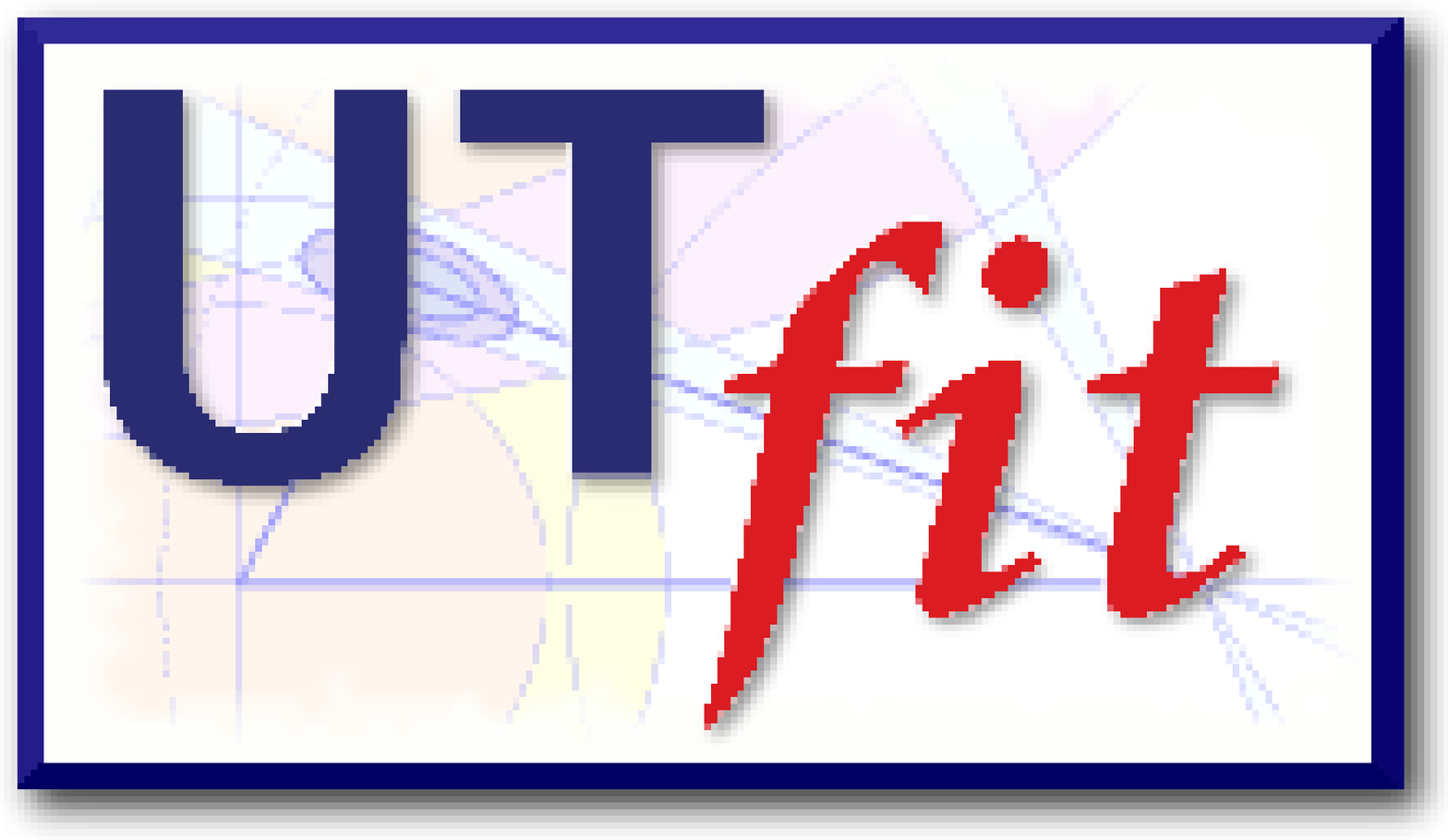,width=2.5cm}
 \end{center}
\end{figure}

\vspace*{-0.8cm}
%\vskip 1. cm
\begin{center}
  \Large{\textbf{UT}}\large{\textit{fit}}\large{~Collaboration} \\
\end{center}
\begin{center}
%  \begin{large}
\vspace*{-0.4cm}
  \textbf{M.~Bona$^{(a)}$, M.~Ciuchini$^{(b)}$, E.~Franco$^{(c)}$,
    V.~Lubicz$^{(b)}$, } \\
  \textbf{G. Martinelli$^{(c)}$, F. Parodi$^{(d)}$, M.
    Pierini$^{(e)}$, P.
    Roudeau$^{(f)}$, }\\
  \textbf{C. Schiavi$^{(d)}$, L.~Silvestrini$^{(c)}$, V.~Sordini$^{(f)}$, A.
    Stocchi$^{(f)}$ and V.~Vagnoni$^{(g)}$}
%  \end{large}
\end{center}
\begin{center}
  \noindent
  \begin{footnotesize}
    \noindent
%    \textbf{$^{(a)}$   INFN,  Sez. di Torino,}\\
%    \hspace*{0.5cm}{Via P. Giuria 1, I-10125  Torino, Italy}\\
    \textbf{$^{(a)}$   Laboratoire d'Annecy-le-Vieux de Physique des Particules,}\\
    \hspace*{0.5cm}{LAPP, IN2P3/CNRS, Universit\'e de Savoie, France}\\
    \textbf{$^{(b)}$ Dip. di Fisica, Universit{\`a} di Roma Tre
      and INFN,  Sez. di Roma Tre,}\\
    \hspace*{0.5cm}{Via della Vasca Navale 84, I-00146 Roma, Italy}\\
    \noindent
    \textbf{$^{(c)}$ Dip. di Fisica, Universit\`a di Roma ``La Sapienza'' and INFN, Sez. di Roma,}\\
    \hspace*{0.5cm}{Piazzale A. Moro 2, 00185 Roma, Italy}\\
    \noindent
    \textbf{$^{(d)}$ Dip. di Fisica, Universit\`a di Genova and INFN, Sez. di Genova,}\\
    \hspace*{0.5cm}{Via Dodecaneso 33, 16146 Genova, Italy}\\
    \noindent
    \textbf{$^{(e)}$ Department of Physics, University of Wisconsin,}\\
    \hspace*{0.5cm}{Madison, WI 53706, USA}\\
    \noindent
    \textbf{$^{(f)}$ Laboratoire de l'Acc\'el\'erateur Lin\'eaire,}\\
    \hspace*{0.5cm}{IN2P3-CNRS et Universit\'e de Paris-Sud, BP 34,
      F-91898 Orsay Cedex, France}\\
%    \textbf{$^{(g)}$ Physik-Department T31, TU-M\"unchen,}\\
%    \hspace*{0.5cm}{D-85748 Garching, Germany}\\
    \textbf{$^{(g)}$INFN, Sez. di Bologna,}\\
    \hspace*{0.5cm}{Via Irnerio 46, I-40126 Bologna, Italy}\\
  \end{footnotesize}
\end{center}

\vspace*{0.5cm}

\begin{abstract}
  Motivated by a recent paper that compares the results of the
  analysis of the CKM angle $\alpha$ in the frequentist and in the
  Bayesian approaches, we have reconsidered the information on the
  hadronic amplitudes, which helps constraining the value of $\alpha$
  in the Standard Model.  We find that the Bayesian method gives
  consistent results irrespective of the parametrisation of the
  hadronic amplitudes and that the results of the frequentist and
  Bayesian approaches are equivalent when comparing meaningful
  probability ranges or confidence levels. We also find that from $B
  \to \pi\pi$ decays alone the 95\% probability region for $\alpha$ is
  the interval $[80^\circ,170^\circ]$, well consistent with recent
  analyses of the unitarity triangle where, by using all the available
  experimental and theoretical information, one gets $\alpha = (93 \pm
  4)^\circ$.  Last but not least, by using simple arguments on the
  hadronic matrix elements, we show that the unphysical region $\alpha
  \sim 0$, present in several experimental analyses, can be
  eliminated.
 \end{abstract}

\newpage \pagestyle{plain}

\section{Introduction}
\label{sec:intro}
Motivated by the criticisms recently appeared in
ref.~\cite{CKMfitteralpha}, we present a new analysis of the CKM angle
$\alpha$ from $B \to \pi \pi$ decays based on the Bayesian statistical
approach. The main results are the following:
\begin{itemize} \item we show that the differences between the
  frequentist and the Bayesian approaches are {\it NOT} due to the
  difference in the two methods but to the difference in the physical
  assumptions on the weak amplitudes, contrary to the claims of
  ref.~\cite{CKMfitteralpha};
\item although we expect an eightfold ambiguity for $\alpha$ using as
  ``a priori'' knowledge only isospin symmetry and the experimental
  measurements of the relevant branching fractions and CP asymmetries,
  {\it this degeneracy can be reduced in the presence of further
    information on the hadronic amplitudes}. We will discuss and use
  this information which can already be extracted from the data;
\item among the information that we do have on the amplitudes, the
  existence of a scale of strong interactions and the use of the
  experimental values of related decays allow to limit the size of the
  hadronic matrix elements and to eliminate the {\it unphysical}
  region of values corresponding to  $\alpha$ close to zero;
\item within our approach, we obtain consistent results at a
  meaningful probability value irrespective of the parametrisation
  used for the hadronic amplitudes.
\end{itemize}
Our analysis is performed within the Standard Model. All the technical
details which are not relevant to the present discussion, but are used
in the analysis, can be found in previous publications of the \utfit{}
Collaboration~\cite{primo,primo1,secondo,ultimo}.

The paper is organised as follows.  We first recall the basic
formalism and definitions of the different quantities entering the
analysis; we then discuss the case in which no further assumptions are
made for these decays besides the experimental knowledge of the
relevant branching fractions and CP asymmetries, and $SU(2)$ isospin
symmetry. In this case we make a comparison between frequentist and
Bayesian methods along the lines of ref.~\cite{CKMfitteralpha}; we
finally discuss the present experimental information on the weak
amplitudes, and the corresponding constraints, and show how this
information helps in reducing the ambiguity of the solution.
 
We think that neither our study nor the study of
ref.~\cite{CKMfitteralpha} can be decisive in settling the long
standing struggle on the validity of the frequentist and Bayesian
methods.  For this reason, contrary to what was done in
ref.~\cite{CKMfitteralpha}, we do not find very useful to present in a
physics paper a long list of learned citations, with philosophical
statements and sentences taken from illustrious references in favour
of the Bayesian statistics. Rather, we thank the authors of
ref.~\cite{CKMfitteralpha} who stimulated us to reconsider the
analysis of the CKM angle $\alpha$, and allowed us to improve our
analysis by using the available experimental information. The
important point is not which of the two approaches is used rather the
ability to get predictions which can eventually be verified ``a
posteriori''. The success of a phenomenological analysis relies on the
capacity of anticipating the correct result.  Among the successful (and
rather accurate) predictions of our collaboration let us recall $\sin
2 \beta$~\cite{primo,pre1} and $\Delta
m_{B_s}$~\cite{primo,primo1,primissimo}, made long before their
experimental measurements.

\section{Generalities}
\label{sec:gen}
The physical quantities which are measured in the experiments and used
in the extraction of $\alpha$ from $B\to \pi \pi$ decays are the
CP-averaged branching fractions ${\cal B}^{+0}_{\pi\pi}$, ${\cal
  B}^{+-}_{\pi\pi}$ and ${\cal B}^{00}_{\pi\pi}$ and the coefficients
${\cal C}^{+-}_{\pi\pi}$, ${\cal C}^{00}_{\pi\pi}$ and ${\cal
  S}^{+-}_{\pi\pi}$of the time-dependent CP asymmetries. At present
${\cal S}^{00}_{\pi\pi}$ is not measured.  The branching fractions and
CP asymmetry coefficients are labelled by the charge of the final state
pions. We give in Table~\ref{tab:expinput} the two sets of values that
we have used in this analysis: SET 1 is the set of
ref.~\cite{CKMfitteralpha}, which will be used for a comparison of the
results; SET 2 is taken from HFAG~\cite{HFAG} and contains the values
updated at the summer 2006 conferences.
%%%%%%%%%%%%%%%%%%%
\begin{table}
\begin{center}
\begin{footnotesize}\begin{tabular}{||c|c|c|c|c|c|c||}
\hline\hline
      &${\cal B}^{+0}_{\pi\pi} $    &   ${\cal B}^{+-}_{\pi\pi}$   &
      ${\cal B}^{00}_{\pi\pi}$  &  ${\cal C}^{+-}_{\pi\pi}$    &
      ${\cal S}^{+-}_{\pi\pi}$ &${\cal C}^{00}_{\pi\pi}$  \\ \hline 
SET 1 ref.~\cite{CKMfitteralpha}&$5.5 \pm 0.6$ &$ 5.1\pm0.4$ &$1.45\pm
0.29$& $-0.37\pm 0.10$ & $-0.50\pm 0.12$ & $-0.28\pm  0.40$ \\ \hline 
SET 2 ref.~\cite{HFAG}&$5.7 \pm 0.4$ &$ 5.2\pm0.2$ &$1.31\pm 0.21$&
$-0.39\pm 0.07$ & $-0.59\pm 0.09$ & $-0.37\pm  0.32$ \\ 
%\\ \hline 
\hline\hline
\end{tabular}\end{footnotesize}
\end{center}
\caption{\it Experimental numbers used in the analysis of the CKM
  angle $\alpha$. The branching fractions are given in units of
  $10^{-6}$.} 
\label{tab:expinput}
\end{table}
%%%%%%%%%%%%%%%%%%%%%
The CP asymmetry coefficients  and branching ratios are related to the
amplitudes by the relations 
\bea  &&{\cal C}^{ij}_{\pi\pi}= (\vert A^{ij}\vert^2 -\vert\bar
A^{ij}\vert^2)/ (\vert A^{ij}\vert^2 +\vert\bar A^{ij}\vert^2)\, , 
\quad \quad   {\cal S}^{ij}_{\pi\pi}= -2 \,{\mathrm Im}[ A^{ij}\bar
A^{ij\, *}]/(\vert A^{ij}\vert^2 +\vert\bar A^{ij}\vert^2) \, ,  \nn
\\ 
&& {\cal B}^{+-,00}_{\pi\pi} =(\vert A^{+-,00}\vert^2 +\vert\bar
A^{+-,00}\vert^2)/2 \, , \quad \quad  {\cal
  B}^{+0}_{\pi\pi}=\tau_{B^+}/\tau_{B^0}\, (\vert A^{+0}\vert^2 +\vert
A^{-0}\vert^2)/2\, . \label{eq:ciu} \eea 
In the equations above we have included in the definition of the
amplitudes trivial and standard factors such as the two body phase
space and the squared Fermi constant.  Here and in the following the
branching fractions are given in units of $10^{-6}$. In our
convention, denoted as ``natural units'' in the following, the
factorised amplitudes are of ${\cal O}(1)$, as explained below.

In the Standard Model, a minimal {\it but non zero} set of ``a
priori'' knowledge on strong interactions is common to all the
phenomenological analyses of $\alpha$. The universal ``a priori''
assumptions of these studies are that strong interactions are flavour
independent and conserve parity and CP. This information is born out
from the experimental measurements of strong interaction processes. In
addition most of the analyses are performed in the approximation in
which isospin symmetry breaking effects - including electromagnetic
corrections - are neglected.~\footnote{For lifetimes we will use the
  experimental value for the charged and neutral $B$
  mesons~\cite{HFAG}.}  In the remaining of this paper we will denote
this set of ``a priori'' assumptions, combined with the isospin
symmetry approximation, ``minimal assumptions''.

Using the ``minimal assumptions'',  in the standard  parametrisation
the amplitudes are written as~\cite{gronau}  
\bea
A^{+-} & \equiv & A(B^0 \to \pi^+ \pi^-) = e^{-i \alpha} \, T^{+-}
\,+\,P\, ,  \nn \\ 
A^{00} & \equiv & A(B^0 \to \pi^0\pi^0) = \frac{1}{\sqrt{2}} \left(
  e^{-i \alpha} \, T^{00} \,-\,P\,  \right) \, ,  \label{eq:sp} \\ 
A^{+0} & \equiv & A(B^+ \to \pi^+\pi^0) = \frac{1}{\sqrt{2}} e^{-i
  \alpha} \, \left( \, T^{00} \,+ \, T^{+-}\,\right) \, . \nn \eea  
The six free parameters are the absolute values of $T^{ij}$ and $P$,
$\vert T^{ij}\vert$ and $\vert P\vert$, their relative strong phases,
$\phi_P=$Arg$[T^{+-} P^*]$ and $\phi_0=$Arg$[T^{+-} T^{00*}]$, the overall phase being
irrelevant, and $\alpha$.

Starting from the general expressions in
Equation~(\ref{eq:sp}), under the ``minimal assumptions'' mentioned
above, one finds for $\alpha$~\cite{gronau}  either zero or eight
solutions from Equations~(\ref{eq:ciu}), corresponding
to~\cite{Charles:1998qx,CKMfitteralpha}
\bea \tan \alpha &=&\frac{ \sin (2 \alpha_{eff} ) \,\bar c + \cos (2
  \alpha_{eff} ) \,\bar s + s}{\cos (2 \alpha_{eff})\,\bar c -\sin (2
  \alpha_{eff})\,\bar s +c}\, ,  \nn \\ 
\sin (2 \alpha_{eff}) &=&\frac{{\cal S}^{+-}_{\pi\pi}}{\sqrt{1-{\cal
      C}^{+-\, 2}_{\pi\pi}}}\, , \quad\quad \cos (2 \alpha_{eff}) =\pm
\sqrt{1-\sin^2 (2 \alpha_{eff}) }\, , \nn \\ 
c &=& \sqrt{\frac{\tau_{B^+}}{\tau_{B^0}}}\,
\frac{\tau_{B^0}/\tau_{B^+} \,{\cal B}^{+0}_{\pi\pi} +{\cal
    B}^{+-}_{\pi\pi} \, \left(1+ {\cal C}^{+-}_{\pi\pi}\right)/2
  -{\cal B}^{00}_{\pi\pi}\left( 1+{\cal
      C}^{00}_{\pi\pi}\right)}{\sqrt{ 2 {\cal B}^{+-}_{\pi\pi}{\cal
      B}^{+0}_{\pi\pi}\left( 1+{\cal C}^{+-}_{\pi\pi}\right)}}
\label{eq:nolabel0}\\  
\bar c &=& \sqrt{\frac{\tau_{B^+}}{\tau_{B^0}}}\,
\frac{\tau_{B^0}/\tau_{B^+} \,{\cal B}^{+0}_{\pi\pi} +{\cal
    B}^{+-}_{\pi\pi} \, \left(1- {\cal C}^{+-}_{\pi\pi}\right)/2
  -{\cal B}^{00}_{\pi\pi}\left( 1-{\cal
      C}^{00}_{\pi\pi}\right)}{\sqrt{ 2 {\cal B}^{+-}_{\pi\pi}{\cal
      B}^{+0}_{\pi\pi}\left( 1-{\cal C}^{+-}_{\pi\pi}\right)}} \nn  
\\ s& = &\pm \sqrt{1-c^2} \, , \quad \quad \bar s = \pm \sqrt{1-\bar
  c^2} \, . \nn \eea 
We also give the formulae from which it is possible to extract, from
the branching fractions and the CP asymmetry coefficients, the
hadronic parameters $T^{ij}$, $P$ and the relative phases
\bea
\vert T^{+-}\vert &=&\left[\frac{{\cal
      B}^{+-}}{2\sin^2\alpha}\left(1\pm\sqrt{1-{\cal C}^{+-\, 2}-{\cal
        S}^{+-\, 2}}\right)\right]^{1/2}\, , \nn \\ 
\vert P\vert &=&\left[\vert
  T^{+-}\vert^2\left(2\cos^2\alpha-1\right)+{\cal
    B}^{+-}\left(1-\frac{{\cal
        S}^{+-}}{\tan\alpha}\right)\right]^{1/2}\, , \nn \\ 
\phi_P&=&\arg(\frac{P}{T^{+-}})=\arctan(x_P,y_P)\, , \nn \\
x_P&=&-\frac{\vert P\vert^2+\vert T^{+-}\vert^2-{\cal
    B}^{+-}}{\cos\alpha}\, , \quad \quad  
%\nn\\
y_P=-\frac{{\cal B}^{+-}\, {\cal C}^{+-}}{\sin\alpha} \, , \label{eq:nolabel}\\
\vert T^{00}\vert&=&\left[ \vert P\vert^2\,\cos 2\alpha+2\,{\cal
    B}^{00}\pm 2\cos^2\alpha\sqrt{\vert P\vert^4-4\,\frac{{\cal
        B}^{00\,2}\,{\cal C}^{00\,2}}{\sin^2 2\alpha}+\frac{\vert
      P\vert^2}{\cos^2\alpha}\left(2{\cal B}^{00}-\vert
      P\vert^2\right)}\right]^{1/2}\, , \nn \\ 
\phi_0&=&\arg(\frac{T^{00}}{T^{+-}})=\phi_P+\arctan(x_0,y_0)\, , \nn \\
x_0&=&\frac{\vert P\vert^2+\vert T^{00}\vert^2-2\,{\cal
    B}^{00}}{2\cos\alpha}\, , \quad \quad  
%\nn \\
y_0=-\frac{{\cal B}^{00}\,{\cal C}^{00}}{\sin\alpha}\, .\nn 
\eea
$\phi_i=\arctan(x_i,y_i)$ is the value of the angle $\phi_i$ obtained
when both $y_i = {\cal F} \sin\phi_i$ and $x_i ={\cal F} \cos \phi_i$
are known (with ${\cal F}$ a positive factor).  In the previous
equation for $\vert T^{+-}\vert$ and $\vert T^{00}\vert$ one must
choose the unique combination of signs that reproduces the correct
result for ${\cal B}^{+0}$

\begin{figure}[ht]
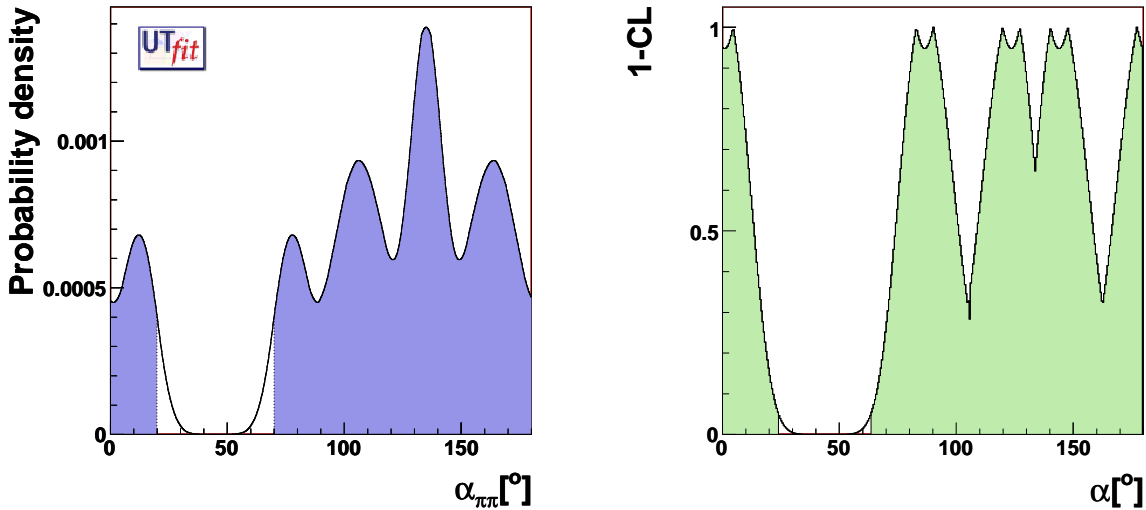

\fone
\caption{{\it Comparison of Bayesian (left) and frequentist (right)
    result in the ``minimal assumptions'' case. The dark band in the
    left figure corresponds to the 95\% probability region, to be
    compared with the 95\% C.L. interval obtained in the right figure.
    The same interval is essentially selected in the two cases. The
    same conclusion holds when comparing the 68\% probability region
    to the 68\% C.L. interval. }}
\label{fig:fig1}
\end{figure}

It has been argued in ref.~\cite{CKMfitteralpha} that using the
``minimal assumptions'' defined above, the frequentist and Bayesian
approaches give different results.  We show that this is not the case
with two clear examples, one based on the present experimental
information, and a second one in the hypothesis that the experimental
errors are reduced by a factor ten.  For this comparison we use the
same data as in ref.~\cite{CKMfitteralpha}, SET 1 of
Table~\ref{tab:expinput}. In Figure~\ref{fig:fig1} we show the
frequentist and Bayesian results with the present experimental
errors.~\footnote{Similar figures were shown in Figure~2 (first and
  fifth) of ref.~\cite{CKMfitteralpha}.}  To obtain these plots, in
the Bayesian approach branching ratios and CP asymmetry coefficients
are extracted with flat a priori probability density function (p.d.f.)
and weighted by the experimental likelihood.  The {\it ES}
parametrisation defined in ref.~\cite{CKMfitteralpha} is the only one
where no priors for the hadronic parameters are specified. Therefore
this is the only case in which no additional physical information with
respect to the frequentist fit is used.
 
Very often there is some confusion in the interpretation of the
results because the shape of the frequentist and Bayesian figures look
very different.  It is important to stress that this happens because
they correspond to two different quantities. Therefore it may be
useful to recall how to read the relevant physical information in the
two cases. Indeed the shape is not the important issue to this
purpose.  In the Bayesian case what it is usually shown is the p.d.f.
for a given physical quantity, whereas in the frequentist case what it
is shown is the confidence level (C.L.), or rather 1-C.L.,
corresponding to a certain value of the physical quantity. Although
there is not a rigorous correspondence, in order to compare the two
approaches one might confront the 68\% (95\%) integrated probability
region of the Bayesian case with the set of values corresponding to
the 68\% (95\%) C.L.\footnote{In our analysis, probability
  regions are delimited by the intercept of the distribution with a
  horizontal line.} In the case of Figure~\ref{fig:fig1} we have
explicitly indicated with a dark band the 95\% integrated probability
region and the region of values corresponding to the 95\% C.L. in the
Bayesian and frequentist case respectively.  The same considerations
apply to the figures shown in the remaining of this paper.

\begin{figure}[ht]
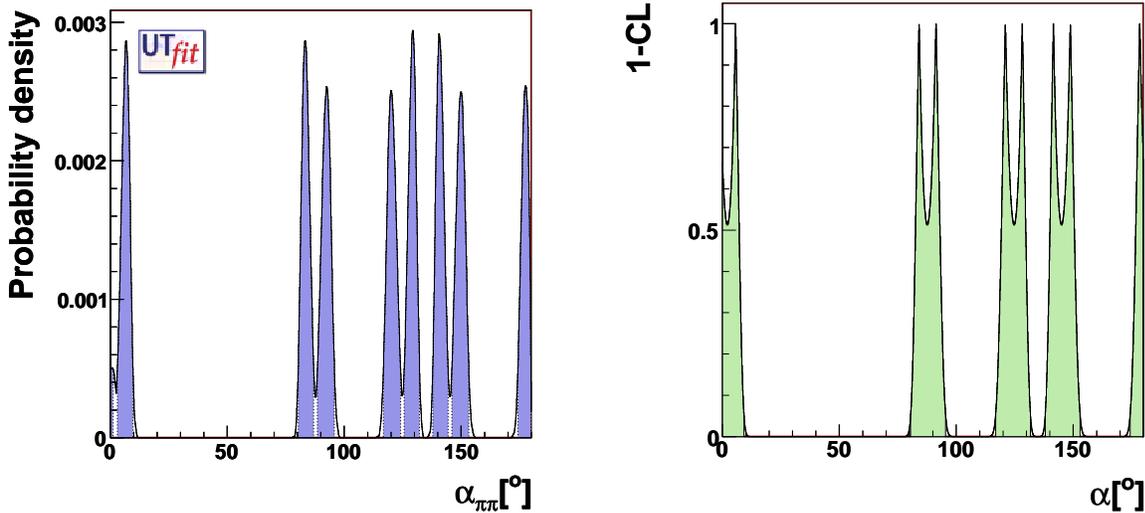

\ftwo
\caption{{\it Same as Figure~\protect\ref{fig:fig1} but with the experimental
    errors reduced by a factor 10. Obviously, also in the Bayesian
    case, an eightfold ambiguity appears.}}
\label{fig:fig2}
\end{figure}

From the selected regions in Figure~\ref{fig:fig1} our conclusion is
indeed just the opposite of what is claimed in~\cite{CKMfitteralpha}:
at a confidence level which is meaningful for obtaining some physical
information, for example at 95\% C.L., and even at the 68\% C.L., when
using the same physical assumptions (in the present case the ``minimal
assumptions''), the frequentist result is equivalent to the Bayesian
expectation, namely the range of values of $\alpha$ between $25^\circ$ and
$65^\circ$ is excluded whereas the complementary interval is fully
allowed. Even in the frequentist case, the eight solutions emerge only
for values of the C.L. smaller than 5\%, which have no physical
meaning.
This is further illustrated by studying the case in which, at fixed
central values, the errors are reduced by a factor of 10: as shown in
Figure~\ref{fig:fig2} the eight solutions are separated both in the
Bayesian and frequentist case. Note that even in the frequentist case
they are still grouped at 95\% C.L.. Thus we
conclude that there is no substantial difference in the physical
information obtained in two approaches, as also discussed in many other 
examples and stated in the Yellow Report of the first CKM
workshop~\cite{1sCKMworkshop}.

\begin{figure}[ht]
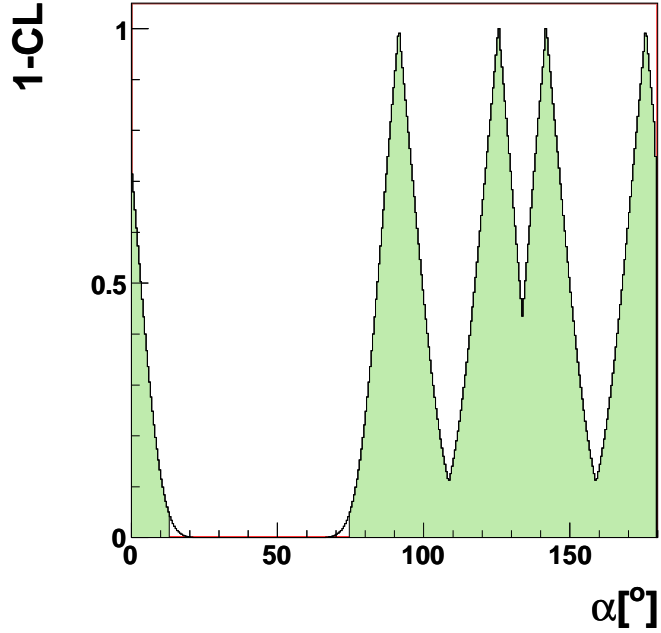

\fsix
\caption{{\it C.L. frequentist plot using the updated experimental
    value from ~\cite{HFAG}.}}
\label{fig:newfreq}
\end{figure}

To strengthen even more the previous arguments we have repeated the
frequentist fit with the updated values corresponding to SET 2 of
Table~\ref{tab:expinput}, taken from HFAG~\cite{HFAG}. The results are
shown in Figure~\ref{fig:newfreq}.  In this case, the new values of
the branching ratios make the eight solutions overlap also in the
frequentist approach (independently of the chosen confidence level).
This shows that the separation of the eight solutions in the
frequentist approach was just a fortuitous accident which may
disappear for small changes in the central values, when the errors are
large.

\begin{figure}[ht]
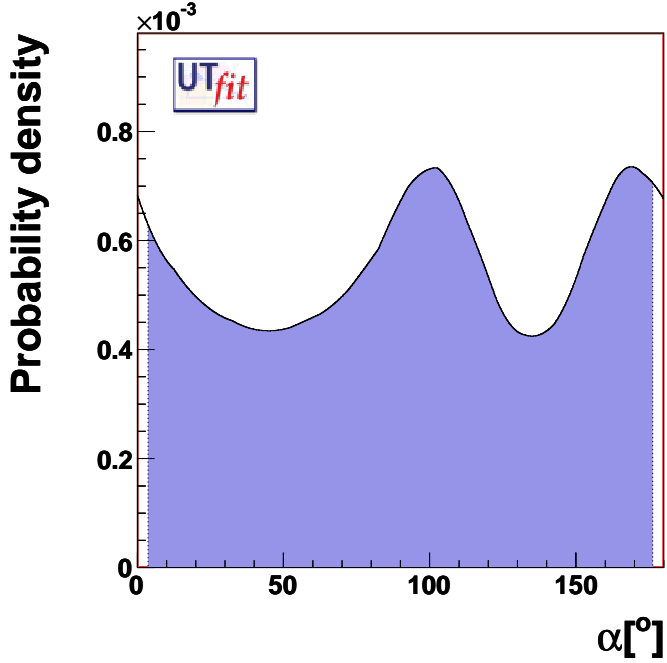

\fthree
\caption{{\it Probability distribution for $\alpha$ obtained removing
    ${\cal B}^{00}_{\pi\pi}$. No significant information can be
    obtained in this case.}}
\label{fig:fig3}
\end{figure}

We have also repeated the exercise of removing some crucial
experimental information, namely we ignored the experimental
measurement of ${\cal B}^{00}_{\pi\pi}$~\cite{CKMfitteralpha}. In the Bayesian approach this
corresponds to let ${\cal B}^{00}_{\pi\pi}$ free to vary between zero
and $10^6$.
% ~\footnote{The authors of ref.~\cite{CKMfitteralpha} blame
%   the Bayesian approach because it was necessary to generate $10^{10}$
%   Monte Carlo events to produce their figures. By using
%   Equations~(\ref{eq:nolabel}), it is straightforward to show that it
%   is possible to obtain accurate distributions with a few hundred
%   thousands events (this is called importance sampling).} 
In this case we get for the p.d.f. of $\alpha$ the result shown in
Figure~\ref{fig:fig3} which allows, for the 95\% total probability
region, almost all the values between $0^\circ$ and $180^\circ$ (corresponding to the  95\% of the total area under the curve), fully consistent with the frequentist approach. We could not reproduce Figure~4 of
ref.~\cite{CKMfitteralpha} (case ES) which we think contains some
error producing the deep at $\alpha \sim 45^\circ$.

For the remaining of this paper, in the analysis we use the more
recent  values of SET 2 in  Table~\ref{tab:expinput}, taken from
HFAG~\cite{HFAG}. 
\section{Information on Hadronic Matrix Elements}
\label{sec:had}
In this Section we present several arguments which show that the size
of the hadronic matrix elements of the operators appearing in the
effective weak Hamiltonian is indeed of the order of magnitude which
is expected in {\it QCD}. This implies that $T^{ij}$ and $P$ are
numbers of ${\cal O}(1)$ in the natural units used in this paper and
that the range of values we used in our previous analyses was too
pessimistic and can be restricted.  With this improvement, as shown in
the next Section, the dependence on the prior noticed in
ref.~\cite{CKMfitteralpha} is substantially eliminated and the
constraint on $\alpha$ improved.

As mentioned before, it is illusory to allege to be able to perform an
analysis of $\alpha$ without a minimal ``a priori'' knowledge about
strong interactions and the hadronic parameters. Such knowledge is
encoded in the expressions given in Equation~(\ref{eq:sp}) which are
valid only if strong interactions are flavour blind and CP conserving,
besides using the approximation that isospin breaking effects are
negligible, which implies $\Lambda_{QCD} \gg (m_d-m_u)$.~\footnote{
  Electroweak penguin contributions and electromagnetic corrections to
  the decay amplitudes are also ignored. See ref.~\cite{IB} for a
  recent discussion of isospin breaking effects in the extraction of
  $\alpha$.}

This however does not exhaust the information that we have on the size
of the matrix elements. A general consideration is that we believe
that the theory of strong interactions is {\it QCD}, which is a
renormalisable theory with a dimensionless coupling constant, and has
a natural scale of ${\cal O}(1\,{\rm GeV})$.  Thus, on dimensional
grounds, in the absence of other scales we expect $\langle M_1
M_2\vert \hat O \vert M \rangle \sim \Lambda_{QCD}^3$ for a
dimension-six four fermion operator $\hat O$.\footnote{For example
  $(\bar b\,u)_{V-A} \, (\bar u \, d)_{V-A}$ in the notation of
  refs.~\cite{buras}.} We would be very surprised to find that this
matrix element has a size of the order of $(1\,{\rm TeV})^3$ or
$M_{Planck}^3$. In the presence of other scales, such as the mass of a
heavy quark, the situation is slightly more complicated, but the
argument is conceptually similar.  Indeed we expect
that~\cite{chernyak}-\cite{SCET0}
\beq
\langle \pi \pi \vert \hat O \vert B \rangle \sim f_\pi M_B^2 f^+(0)
\sim f_\pi M_B^2 \left(\frac{\Lambda_{QCD}}{M_B}\right)^{3/2} \sim
M_B^{1/2} \Lambda_{QCD}^{5/2} \, . \label{eq:size}
\eeq 
Note that this
simple dimensional argument, which we have derived using factorised
expressions in the intermediate steps, has a more general validity
than factorisation although, obviously, factorisation respects the
same scaling laws. For chirally enhanced contributions, the
enhancement factor, by which the expression in Equation~(\ref{eq:size})
must be multiplied, is  of the order $ 2
M_\pi^2/(m_u+m_d)/M_B\sim 0.8$ and thus it does not change the natural
size of the amplitude.  It is straightforward to show, see
Equation~(\ref{eq:estfact}) below, that this implies $T^{ij} \sim 1$.  Again
we would be very surprised to find that, in order to reproduce the
experimental branching ratios, the size of the matrix elements must be
much larger than its natural size. In the  language  of factorisation this
would correspond to dimensionless $B$-parameters much larger than one.

We now provide further support to these general considerations:
\begin{enumerate} 
\item The first exercise, useful to estimate the range of values which
  can be expected, is to compute $T^{+-}$ using only current-current
  operators and strict factorisation, namely
  \bea  \vert T^{+-}\vert^2&=&
  % BR(B^0_d \to \pi^+ \pi^-) 
  % \frac{32 \pi M_B}{G_F^2\,\vert V_{ub}V_{ud}^*\vert^2\,\tau_B}
  % =\nn \\
  \frac{ G_F^2\tau_B  \vert V_{ub}V_{ud}^*\vert^2}{32 \pi M_B} 
  \vert C_1(M_B) \langle \pi^+\pi^-\vert O_1\vert B^0_d\rangle
  +C_2(M_B) \langle \pi^+\pi^-\vert O_2\vert B^0_d\rangle\vert^2
  \times 10^6 \nonumber  \\ &=& 
  \frac{ G_F^2\tau_B  \vert V_{ub}V_{ud}^*\vert^2}{32 \pi M_B} 
  \left\vert\frac{ C_1(M_B)}{3}+ C_2(M_B) \right\vert^2  \times  \vert
  M_B^2 f_\pi f^+(0)\vert^2 \times 10^6\,.\label{eq:estfact} \eea 
  Using $f_\pi=132$~MeV, $f^+(0)=0.3$, $C_1(M_B)=-0.2$ and
  $C_2(M_B)=1.1$, we find $\vert T^{+-}\vert=3.2$ in natural units. 

\item Long before it was experimentally measured, several predictions
  based on factorisation or other approaches existed for ${\cal
    B}^{+-}_{\pi\pi}$, ${\cal B}^{00}_{\pi\pi}$ and ${\cal
    B}^{+0}_{\pi\pi}$.  The latter is the simplest to predict since it
  only depends on emission diagrams, without penguins, annihilations
  or further complications. In Table~\ref{tab:predexp} we compare some
  predictions, obtained in different theoretical frameworks, with the
  present experimental determination.  We find that the order of
  magnitude of the predicted branching fractions, based on values of
  the hadronic amplitudes of ${\cal O}(1)$ in natural units, are
  approximately in agreement with the experimental value.
  %%%%%%%%%%%%%%%%%%% 
  \begin{table}
    \begin{center}
      \begin{tabular}{||c|c|c|c||}
        \hline\hline  
        ref.~\cite{Ciuchini:1997rj}
        & ref.~\cite{bbns} & ref.~\cite{Keum:2002cr} & Exp. \\
        $3.6-5.3 $ & $4.3\,(1\pm 0.3)$ & $3.7^{+1.3}_{-1.1}$ & $5.5\pm 0.6$\\
        \hline\hline
      \end{tabular}
    \end{center}
    \caption{\it Comparison of the experimental value of ${\cal
        B}^{+0}_{\pi\pi}$ with  the theoretical
      predictions \cite{Ciuchini:1997rj,bbns,Keum:2002cr}. All values
      are given in units of
      $10^{-6}$.} 
    \label{tab:predexp}
  \end{table}
  %%%%%%%%%%%%%%%%%%%%% 
%   Obviously we cannot exclude that all these theoretical estimates are
%   wrong and that each of the two emission diagrams (alternatively each
%   of the two amplitudes $T^{+-}$ and $T^{00}$) gives a much larger
%   contribution which cancels out with the other to mimic an amplitude of
%   ${\cal O}(1)$.  However this would be really peculiar and contrary to
%   the general physical argument, supported by a plethora of examples, on
%   the presence of a typical scale in {\it QCD}.
\item The third argument in favour of the natural order of magnitude
  for the matrix elements is based on the scaling of the rates between
  $B$ and $D$ decays.  Since, in the heavy quark limit, the partial
  decay rates scale as the squared amplitude divided the heavy mass,
  namely as $ 1/M \times \vert M^{1/2} \Lambda_{QCD}^{5/2}\vert^2 \sim
  \Lambda_{QCD}^{5/2}$, if we take the ratio 
  \beq R = \frac{\vert
    T^{+-}(B^0_d \to \pi^+ \pi^-)\vert^2}{\vert T^{+-}(D^0 \to \pi^+
    \pi^-)\vert^2} \sim \frac{\vert V_{ub}V_{ud}^*\vert^2}{\vert
    V_{cd} V_{ud}^*\vert^2} \, ,
  \eeq 
  this is independent of the meson
  mass and can be used to extract the absolute value of the unknown
  hadronic parameter $T^{+-}$ as follows 
  \beq \vert T^{+-}\vert^2 =
  BR(D^0 \to \pi^+ \pi^-)\times 10^6 \,\frac{\tau_{B^0_d} }{\tau_{D^0}
  } \, R \, . 
  \eeq
  Using the central value of the  experimental measurement $BR(D^0 \to \pi^+
  \pi^-) = 1.5 \times 10^{-3}$, $\tau_D= 0.41 \times 10^{-12}$~sec,
  $\tau_B= 1.6 \times 10^{-12}$~sec, $\vert V_{ub}\vert= 3.7 \times
  10^{-3}$, $\vert V_{cd}\vert=0.22$, we find $\vert T^{+-}\vert=1.3
  $.  
%   We emphasize again that there is the possibility, rather remote
%   in our opinion, that factorization is broken by some order of
%   magnitude, that $T^{+-}$ and $P$ are much larger than one, and that
%   every contribution conspires to cancel out in order to mimic much
%   smaller amplitudes, of the expected size.
\item We can get some knowledge about the parameter $P$ from the study of   $B_s
  \to K^+ K^-$ decay. Up to doubly Cabibbo suppressed terms, this
  decay proceeds only through the penguin contribution $P_s$, which
  corresponds to $P$ up to $SU(3)$ breaking
  effects~\cite{burassilvestrini}.  Using the central value of the
  experimental measurement $BR(B_s \to K^+ K^-)= (24.4 \pm 1.4 \pm
  4.6) \times 10^{-6}$~\cite{ckmpunzi}, and the relation 
  \beq \vert
  P\vert^2 = BR(B_s \to K^+ K^-)\times 10^6 \,\frac{\tau_{B^0_d}
  }{\tau_{B^0_s} } \,\ \frac{\vert V_{td}V_{tb}^*\vert^2}{\vert
    V_{ts}V_{tb}^*\vert^2} \, , \eeq 
  we find $\vert P_s\vert = 1.1$.
  Even accepting the very pessimistic point of view that $SU(3)$
  breaking effects are 100\%, one still obtains a number of 
  ${\cal O}(1)$  which
  automatically constrains, when combined with $B \to \pi\pi$ decays,
  also $T^{+-}$ to be of ${\cal O}(1)$.  For $B_s \to K^+ K^-$ this argument
  could be spoilt if the Cabibbo suppressed emission amplitude, $T$,
  were much larger than the penguin amplitude, namely if $\vert T\vert
  \gg \vert P_s \vert \gg 1$ in natural units. This possibility is
  however excluded because, in order to reproduce the experimental
  values of ${\cal B}^{+-}_{\pi\pi}$ and ${\cal B}^{00}_{\pi\pi}$, we
  may have $\vert T \vert \gg 1$ and $\vert P\vert \gg 1$ but it is
  necessary that $\vert T \vert \sim \vert P\vert $.
\end{enumerate}
In summary all the experimental and theoretical evidence is in favour
of matrix elements of ${\cal O}(1)$ in ``natural units''.  The
arguments from 1. to 3. show that in our previous
analyses~\cite{secondo,ultimo} the range that we used by varying
$\vert T^{ij}\vert $ and $\vert P\vert$ between $0$ and $10$ was
rather conservative and easily accommodates $\mathcal{O}(1)$
corrections to the theoretical estimates presented above. Argument 4.
puts a stronger constraint on the penguin contribution since, even
assuming an $SU(3)$ breaking effect as large as 100\%, it is difficult
to imagine that it could exceed the value $\vert P\vert =2.5$ in
natural units.  As discussed in 4.  this upper value on $\vert P\vert$
limits the acceptable values of $\vert T^{+-,00} \vert$.

We conclude this Section with a few comments on appendix B of
ref.~\cite{CKMfitteralpha}.  In this appendix, the authors criticise
the Bayesian approach for being unable to reproduce the peak of the
p.d.f. at $\alpha$ around zero.  We believe instead that they are
completely mislead by their prejudice of ignoring the existing
information on the Standard Model and on the hadronic matrix elements
and for this reasons they are unable to eliminate the unphysical
solutions at $\alpha$ close to zero, which cannot be
there.~\footnote{Furthermore they find the rather original result that
  the C.L. is about 90\% for $\alpha \sim 10^{-30}$ and zero for
  $\alpha=0$!!}  A very simple argument kills simultaneously the
solutions at $\alpha \sim 0$ and their arguments to show that the
Bayesian approach is unable to cope with the RI parametrisation, where
$T^{+-}=\tau^{+-}/\sin\alpha$ and the prior distribution is taken for
$\tau^{+-}$ rather than for $T^{+-}$.  For example, it is
straightforward to show that $\alpha < 2^\circ$ corresponds to $T^{+-}
> 30$ in natural units. This implies that, in order to fit the
branching ratios, also $P \sim 30$, i.e. an $SU(3)$ breaking effect
of about 3000\%!  Bearing in mind that $m_s/m_d \sim 10$ we would then
expect $SU(2)$ effects of ``only'' 300\% which should, in our
subjective and humble opinion, be taken into account invalidating the
assumption of neglecting  isospin breaking effects in the analysis.

\section{Adding Useful Information}
\label{sec:adding}
In Section~\ref{sec:gen} we have shown that, using the ``minimal
assumptions'' on the hadronic matrix elements,  no real difference in
the physical information at a significant level of confidence exists
between the Bayesian and the frequentist approach; in
Section~\ref{sec:had} we have discussed with several examples the
information that we have on the weak hadronic matrix elements. All the
arguments support the existence of a typical size for the hadronic
matrix elements, which constrains the range of possible values.  We do
not understand why one should ignore this knowledge while using the
simplification (and the information) that comes from isospin symmetry.

In this Section, we make use of the constraints on the size of the
matrix elements in different ways.  In general, we show that the
information on the matrix elements helps eliminating some of the eight
solutions that exist in the ``minimal assumptions'' case.  The
prejudice that any acceptable method must lead to  eight
solutions~\cite{CKMfitteralpha} is obviously wrong: if we could
determine exactly the absolute values of the hadronic matrix elements
and the strong phases from a theoretical calculation, from the
experimental values of ${\cal B}^{ij}_{\pi\pi}$ and of the CP
asymmetry coefficients we could extract unambiguously the value of $
\alpha$.  A partial knowledge, or some constraint, will then remove at
least partially the degeneracy of the solutions.

In particular, we show that, using the constraints on the hadronic
matrix elements, not only we eliminate the pathological solution at
$\alpha \sim 0$, which could survive only with stratospheric values of
the matrix elements, but we can substantially reduce the difference
between the different parametrisations that was emphasised in
ref.~\cite{CKMfitteralpha}. Indeed the difference disappears for the
95\% probability region, and is marginal even for the 68\% one.

For the parametrisations we use the same convention as in
ref.~\cite{CKMfitteralpha}: {\it MA} in which the absolute values of
the hadronic amplitudes and their phases are extracted with a flat
p.d.f.; {\it RI} in which the real and imaginary part of the hadronic
matrix elements are extracted with a flat p.d.f.; {\it ES} in which,
in the Bayesian case, we generate the measured quantities, namely
${\cal B}^{+0}_{\pi\pi}$, ${\cal B}^{+-}_{\pi\pi}$ and ${\cal
  B}^{00}_{\pi\pi}$ and the CP asymmetry coefficients ${\cal
  C}^{+-}_{\pi\pi}$, ${\cal C}^{00}_{\pi\pi}$ and ${\cal
  S}^{+-}_{\pi\pi}$, with flat a priori probability density function
(p.d.f.) and weighted with the experimental likelihood. In this case,
when solving for the absolute value of the matrix elements according
to Equations~(\ref{eq:nolabel}) of Section~\ref{sec:gen}, we only accept
the solutions when they fall in the ``a priori'' acceptable region
defined below.

For the following discussion, we distinguish the ``previous'' set of a
priori, which corresponds to the ranges used in our previous analyses
($\vert T^{ij}\vert \le 10$, $\vert P\vert \le 10$, and arbitrary
phases)~\cite{secondo} and the ``current'' one, corresponding to
$\vert T^{ij}\vert \le 10$, $\vert P\vert \le 2.5$ and arbitrary
phases. In the last case we limit $\vert P\vert$ to be less than $2.5$
since, to our knowledge, no physical quantity suffers from a $SU(3)$
breaking effect larger than 100\% and from $B^0_s \to K^+ K^-$ we
estimated $\vert P\vert \sim 1.1$.  In this Section we always use the
values of the measured quantities corresponding to SET 2 in
Table~\ref{tab:expinput}.
\begin{figure}[ht]
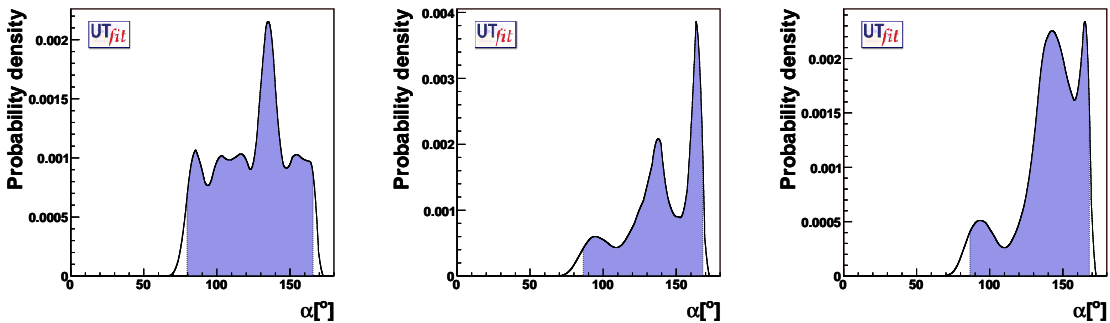

\ffour
\caption{{\it Probability distribution for $\alpha$ obtained with
    three different prior assumptions: {\it ES} (left), {\it MA}
    (centre) and {\it RI} (right). In all the cases the hadronic parameter
    are constrained by $\vert T^{ij}\vert \le 10$, $\vert P\vert \le
    2.5$. In all the cases the region with ''low'' $\alpha$ is excluded.
    This is not due to the statistical approach, but to the physical
    assumptions on the hadronic matrix elements (the smooth aspect of
    these p.d.f.s is due to the fact we use only $10^5$ (sic!) Monte
    Carlo events supplemented with an efficient numerical
    algorithm).}}
\label{fig:comparison}
\end{figure}

In Figure~\ref{fig:comparison} we show the probability distribution
functions for the three parametrisations ({\it MA}, {\it RI} and {\it
  ES}), in the ``current'' case.  A comparison of the 95\% probability
region, highlighted in dark, show that the different parametrisations
give the same physical information, although the shape of the p.d.f.s
looks different.  The difference observed in
ref.~\cite{CKMfitteralpha} between {\it MA} and {\it RI} in the
``previous'' case was due to the different ``a priori'' distribution
for the absolute values of the amplitudes ({\it RI} has a p.d.f.
linearly growing with the absolute value) and the lack of further
information on the hadronic matrix elements.  In this respect we have
to admit that we were not Bayesian enough in our previous analyses,
because the ``previous'' upper limit consisted in taking the
``intuitive'' upper bound one order of magnitude larger than the
typical size, without using the experimental information which was
already available.

Indeed the upper bound on $P$, which was obtained using only $SU(3)$
arguments, implies a bound also on $\vert T^{ij}\vert$.  If $P$ is
limited, in order to reproduce the observed $B\to \pi\pi$ branching
ratios, $\vert T^{ij}\vert$ cannot be too large.  We leave to our
friends of ref.~\cite{CKMfitteralpha} the exercise to show that with
the same physical inputs one gets in the frequentist case the same
results at the 95\% C.L..

After the above discussion, where we have shown that the Bayesian
approach gives physically meaningful results, we are now ready to
discuss the implications of the bound that we have imposed on the size
of the hadronic amplitudes, in particular the penguin amplitude $\vert
P \vert $.

\begin{figure}[t!]
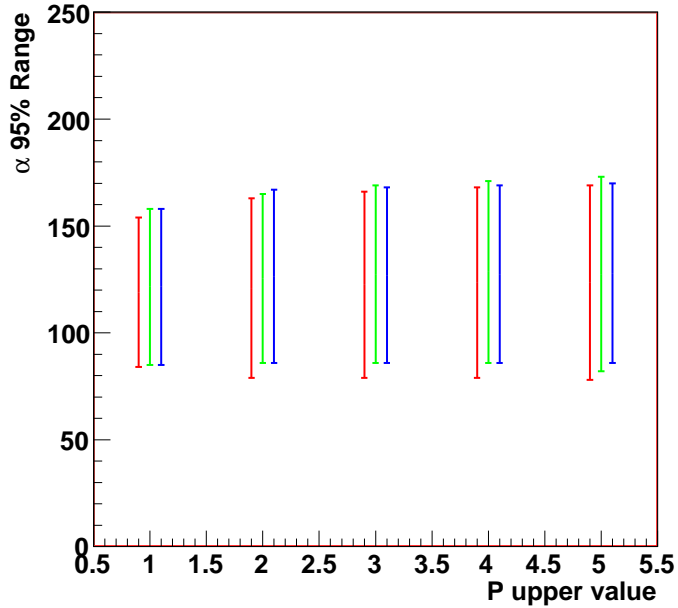

\ffive
\caption{{\it We show the range of allowed values of $\alpha$ corresponding to
     the 95\% probability regions for the three parametrizations
    ({\it MA}=red, {\it RI}=green, {\it ES}=blue) as a function of the
    maximum allowed value of $\vert P\vert$, $P_{MAX}$, scanning  the range     $1 \le 
    P_{MAX} \le 5.5$. This figure shows that the residual dependence on
    $\vert P\vert$ is very mild.}}
\label{fig:variation}
\end{figure}

Even in the case of observed CP violation in $B \to \pi\pi$ decays,
the system of equations used to extract $\alpha$ and the hadronic
amplitudes, including the strong phases, from branching fractions and
CP asymmetry coefficients, Equations~(\ref{eq:nolabel0}) and
(\ref{eq:nolabel}), admits the unphysical solution $\alpha=0$.  This
solution can only be obtained however in the peculiar limit $\vert T
\vert \to \infty$, $ \vert P\vert \to \infty$ and $P /T \to -1$. This
has several unappealing features: i) in the Standard Model, CP
violation must disappear for $\alpha \to 0$ since the unitarity
triangle collapses to a line and the Jarlskog determinant vanishes;
ii) the hadronic amplitudes cannot go to infinity without violating
the basic properties of any renormalisable field theory (and we have
discussed what is their expected size);~\footnote{In addition to the
  arguments given in the text, the limit mentioned above would require
  an infinite amount of fine tuning and correlation among parameters
  related to different physics, namely $\alpha$ which has an
  electroweak interaction origin and the hadronic amplitudes which are
  sensitive to strong interactions.} iii) the unphysical solutions for
$\alpha \sim 0$ are eliminated, since we have shown that they imply
arbitrary large $SU(3)$ breaking effects, which are not consistent
with the assumption of isospin symmetry in the expressions of the
amplitudes.

\begin{figure}[t!]
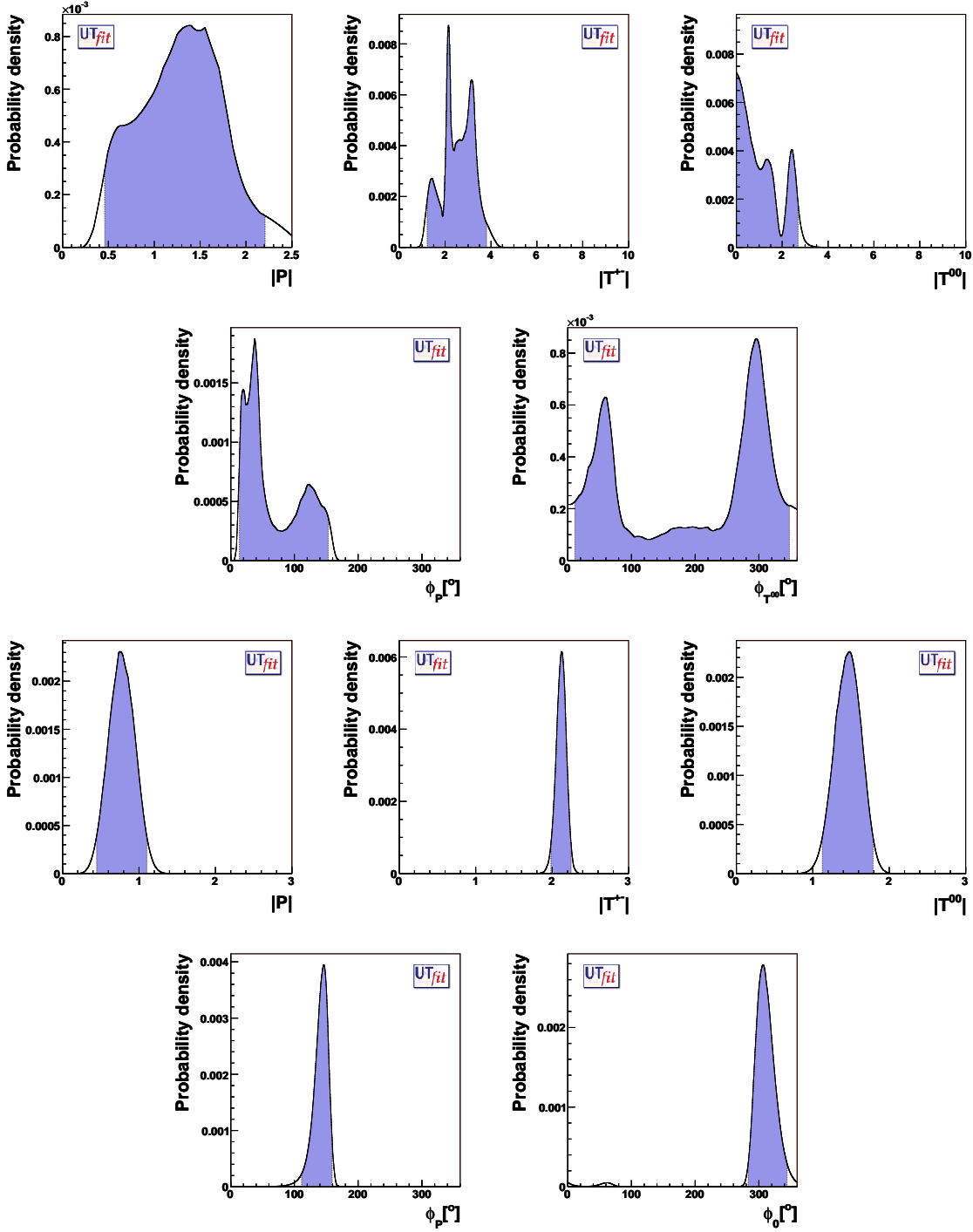

\fseven
\caption{{\it We show the p.d.f.s for $\vert P \vert$, $\vert T^{+-}
    \vert $ and $\vert T^{00} \vert $, and the relative phases,
    obtained in the {\it MA} parametrisation, using only $B \to
    \pi\pi$ decays (first two rows) or the full UT analysis (last
    two rows).}}
\label{fig:pdfs}
\end{figure}

The bounds on the hadronic amplitudes discussed in
Section~\ref{sec:had} allow to get rid of the unphysical solution in a
straightforward way.  Rather than living with a non physical solution,
we prefer by far to remain with the smallish residual dependence on
the upper value on $\vert P \vert$ on the allowed interval for
$\alpha$, shown in Figure~\ref{fig:variation}. It would be interesting
to compare the uncertainty originating from the upper value on $\vert
P \vert$ to the size of the uncertainty due to the corrections coming
from the neglected isospin breaking effects.

For completeness, we finally give in Figure~\ref{fig:pdfs} the p.d.f.s
for the relevant amplitudes and strong phases, as selected by our
analysis, in the {\it MA} parametrisation. These distributions can be
used for comparison and test of the different theoretical approaches
which compute, within some approximation, the hadronic matrix elements
from {\it QCD} ~\cite{bbns,Keum:2002cr,SCET,hep-ph/0506228,hep-ph/0404073}.
We present two cases: the case in which only the quantities measured
in $B \to \pi\pi$ decays are used, and the case where, in the Standard
Model, we make a combined CKM analysis using all the available
experimental and theoretical information~\cite{primo,primo1,secondo,ultimo}.
The results confirm the a priori knowledge on the weak hadronic matrix
elements based on dimensional analysis and the existence of a typical
scale in {\it QCD}, namely the allowed values of $\vert
T^{+-,00}\vert$ and $\vert P\vert$ are of ${\cal O}(1)$.  To be more
precise, in the full Standard Model analysis, from the 68\% region, we
find $\vert P \vert = 0.80\pm0.24$, $\vert T^{+-} \vert = 2.1 \pm0.1$
and $\vert T^{00} \vert = 1.4 \pm0.2$, in full agreement with the
expectations based on simple physical arguments of {\it QCD} and
discussed in Section~\ref{sec:had}.

\section*{Conclusions}
Stimulated by a recent paper on the extraction of the CKM angle
$\alpha$ from $B \to \pi\pi$ decays~\cite{CKMfitteralpha}, we have
upgraded our Bayesian analysis with the following results: i) we have
shown that the present information on the hadronic matrix elements,
obtained from general theoretical arguments and experimental
measurements, already allows a substantial reduction of the eightfold
ambiguity in the determination of $\alpha$, in particular by
eliminating the solutions at $\alpha \sim 0$, that correspond to
unphysical values of the amplitudes; ii) the information on the
hadronic matrix elements substantially eliminates, in the Bayesian
approach, the dependence of the results on the ``a priori'' probability
distributions of the hadronic matrix elements, which was noticed in
ref.~\cite{CKMfitteralpha}. 
Contrary to the claims of ref.~\cite{CKMfitteralpha}, we have also shown that the differences between the frequentist and
the Bayesian approaches are not due to the difference in the two
methods but to the difference in the physical assumptions on the weak
amplitudes. We believe that a continuation of this sterile polemics in
favour of the frequentist or Bayesian  approach is only a waste of
time and energies and it is better to concentrate the efforts in
trying to combine in the most efficient way the rich information which
is coming from the measurements of several non-leptonic channels to
constrain the CKM parameters and to make accurate predictions.

\section*{Acknowledgements}
This work has been supported in part by the EU networks ``The quest for
unification'' under the contract MRTN-CT-2004-503369 and
``FLAVIAnet'' under the contract MRTN-CT-2006-035482.

\end{document}